\RequirePackage{silence}
\documentclass[11pt,a4paper,twoside]{article}
\usepackage{cmap}
\usepackage[utf8]{inputenc}
\usepackage[T1,T2A]{fontenc}
\usepackage[english.ancient,russian.ancient]{babel}
\usepackage{graphicx}
\usepackage{array}
\usepackage[doi]{samgtu-bib}
\usepackage{samgtu}
\usepackage{mathtools}
\mathtoolsset{showonlyrefs}
\usepackage{desclist}
\usepackage{euscript}
\usepackage{mathrsfs}
\usepackage{multicol}
\usepackage{mathrsfs}
\usepackage{cite}
\usepackage{qrcode}
\allowdisplaybreaks[4]
\usepackage[nodayofweek]{datetime}

\renewcommand{\JournalYear}{2023}
\renewcommand{\JournalVolume}{xx}
\renewcommand{\JournalNumber}{x}

\graphicspath{{./00PIC/}} 

\vsgtu{xxxx} 
\FirstPage{1}
\LastPage{x}

\pdfcrop

\begin{document}

\udc{519.635}
\msc{65M06}

\rutitle{Подавление пилообразных осцилляций при использовании разностной схемы для моделирования массопереноса в высыхающей на подложке капле в приближении тонкого слоя}
\rutitleshort{Подавление пилообразных осцилляций}

\entitle{Suppression of sawtooth oscillations when using a finite-difference scheme for mass transfer simulation via the lubrication approximation in a droplet evaporated on a substrate}
\entitleshort{Suppression of sawtooth oscillations}

\ruauthor{К.~С.~Колегов\affil{1,2}}{К\,о\,л\,е\,г\,о\,в~К.~С.}
\enauthor{K.~S.~Kolegov\affil{1,2}}{K\,o\,l\,e\,g\,o\,v~K.~S.}

\ruaffil{Астраханский государственный университет имени В.~Н.~Татищева,\\
Россия, 414056 , Астрахань, ул. Татищева, 20а.
\and
Тюменский государственный университет, \\
Россия, 625003, г. Тюмень, ул. Ленина, д. 25.}

\enaffil{Astrakhan State University named after V.N. Tatishchev, \\
20a, Tatischev st., Astrakhan , 414056, Russian Federation.
\and
University of Tyumen,\\
25, Lenin st., Tyumen, 625003, Russian Federation.}

\ruauthorsdetails{2}{
\fullauthor{Колегов Константин Сергеевич}
\orcid{0000-0002-9742-1308} 
\regalia{кандидат физико-математических наук}
\position{старший научный сотрудник}
\dept{лаб. <<Математическое моделирование и информационные технологии в науке и образовании>>}
\email{konstantin.kolegov@asu.edu.ru}
}

\enauthorsdetails{2}{
\fullauthor{Konstantin S. Kolegov}
\orcid{0000-0002-9742-1308} 
\regalia{Cand. Phys. \& Math. Sci.}
\position{Senior researcher}
\dept{Lab. of Mathematical Modeling}
\email[br]{konstantin.kolegov@asu.edu.ru}
}

\rukeywords{испаряющаяся капля, массоперенос, капиллярный поток,  разностная схема, пилообразная осцилляция, подавление эффекта коффейных колец}
\enkeywords{evaporating droplet, mass transfer, capillary flow, finite-difference scheme, sawtooth oscillation, suppression of the coffee-ring effect}

\ruabstract{Испаряющиеся капли и плёнки используются в приложениях из разных областей. Особый интерес представляют различные методы испарительной самосборки. В работе описана математическая модель массопереноса в высыхающей на подложке капле на базе приближения тонкого слоя. Модель учитывает перенос растворённого или взвешенного вещества капиллярным потоком, диффузию этого вещества, испарение жидкости, формирование твёрдого осадка, зависимость вязкости и плотности потока пара от концентрации примеси. Рассматривается случай, когда трёхфазная граница <<жидкость--подложка--воздух>> закреплена. Для уравнений модели разработаны явные и неявные разностные схемы. Предложена модификация численного метода, в которой комбинируется расщепление по физическим процессам, итерационный метод явной релаксации и метод прогонки. Описан практический рецепт подавления пилообразных осцилляций на примере конкретной задачи. Разработан программный модуль на языке С++, который в дальнейшем можно будет использовать для задач испарительной литографии. С помощью этого модуля проведены численные расчёты, результаты которых сравнивались с результатами, полученными в пакете Maple. Численное моделирование предсказало случай, когда направление капиллярного потока с течением времени меняется на противоположное из-за изменения знака градиента плотности потока пара. Это может приводить к замедлению выноса вещества на периферию, что в результате будет способствовать формированию более или менее равномерного осадка по всей площади контакта капли с подложкой. Данное наблюдение полезно для совершенствования методов подавления кольцевых осадков, связанных с эффектом кофейных колец и нежелательных для некоторых приложений, как, например, струйная печать или нанесение покрытий.}

\enabstract{Evaporating droplets and films are used in applications from different fields.  Various methods of evaporative self-assembly  are of particular interest. The paper describes a mathematical model of mass transfer in a droplet drying on a substrate based on the lubrication approximation. The model takes into account the transfer of a dissolved or suspended substance by a capillary flow, the diffusion of this substance, the evaporation of liquid, the formation of solid deposit, the dependence of the viscosity and the vapor flux density  on the admixture concentration. The case with pinning of the three-phase boundary (``liquid--substrate--air'') is considered  here. Explicit and implicit finite-difference schemes have been developed for the model equations. A modification of the numerical method is proposed, in which splitting by physical processes, the iterative method of explicit relaxation and Thomas algorithm are combined. A practical recipe for suppressing sawtooth oscillations is described using the example of a specific problem. A software module in C++ has been developed, which can be used for evaporative lithography problems in the future. With the help of this module, numerical calculations were carried out, the results of which were compared with the results obtained in the Maple package. Numerical simulation predicted the case in which the direction of the capillary flow changes to the opposite over time due to a change in the sign of the gradient of the vapor flux density. This can lead to a slowdown in the transfer of the substance to the periphery, which as a result will contribute to the formation of a more or less uniform precipitation over the entire contact area of the droplet with the substrate. This observation is useful for improving methods of annular deposit suppression  associated with the coffee-ring effect and undesirable for some applications, such as inkjet printing or coating.}


\makerutitle


\Section[n]{Введение}
Испаряющиеся капли и плёнки используются в приложениях из разных областей, например, охлаждение нагретых поверхностей электронных приборов, диагностика в медицине, формирование прозрачных электропроводных покрытий на гибкой подложке, структурирование поверхности~\cite{Zang2019,Kolegov2020}. Метод испарительной литографии появился после выяснения связи возникающего при испарении капель коллоидных растворов эффекта кофейных колец~\cite{Deegan1997} с естественным образом формирующимися неоднородными потоками пара с поверхности капли (см. обзор~\cite{Kolegov2020}). В методе испарительной литографии контролируемое создание пространственных структур в осадках, остающихся на подложке после высыхания жидкости, достигается при помощи внешних условий, индуцирующих неравномерное испарение с поверхности коллоидной жидкости. Осадки могут оставаться не только на подложке, но и на стенке ячейки~\cite{Kim2022}. Испарительная литография является частью более широкого направления. Речь идет о самоорганизации, вызванной испарением (evaporative-induced self-assembly (EISA)). К этому обширному направлению относятся методы на основе процессов, связанных с контактной линией (граница трех фаз <<жидкость--подложка--воздух>>), методы на основе сил межчастичного взаимодействия и испарительная литография. Как правило, испарительная литография является гибким и одноступенчатым процессом, преимущества которого связаны с простотой, дешевизной и применимостью практически к любой подложке без предварительной обработки. В такой литографии отсутствует механическое воздействие на шаблон, поэтому его целостность в процессе работы не нарушается. Также этот метод полезен для создания материалов с локализованными функциями, такими как скользкость и самовосстановление. По этим причинам испарительная литография привлекает все большее внимание и к настоящему времени имеет ряд достижений. В~\cite{Kolegov2020} также проанализированы имеющиеся ограничения рассматриваемого метода и пути его дальнейшего развития.
	
	 Методы испарительной литографии можно разделить на активные и пассивные. Их отличие в том, что первая подгруппа характеризуется наличием ключевых параметров, которые регулируются в режиме реального времени, а вторая подгруппа подразумевает наличие статических ключевых параметров, которые настраиваются до начала процесса~\cite{Kolegov2020}. К отдельной подгруппе относятся гибридные методы, которые также могут быть как пассивными, так и активными. Эти методы сочетают в себе испарительную литографию с другими методами, относящимися и не относящимися к испарительной самосборке~\cite{Kolegov2020}. Испарительная литография предоставляет больше возможностей для формирования структур различной геометрической формы на микро- и наноуровне по сравнению с испарительной самосборкой. Но с другой стороны испарительная самосборка позволяет получать структуры меньшие по размеру. Разработка новых гибридных методов в испарительной литографии позволит получать структуры осадков или рельефных твердых пленок требуемой формы и морфологии еще меньших размеров для относительно больших площадей. Такие покрытия будут устойчивы к внешним механическим воздействиям, в них будут отсутствовать трещины~\cite{Baba2021}. Кроме того, эти покрытия можно будет наделить некоторыми требуемыми функциональными свойствами.
	
	 К примеру, для разработки гибридных методов можно использовать дополнительное воздействие через различные факторы: пропускание электрического тока через подложку с испаряющейся каплей~\cite{Wang2020}, влияние магнитным полем на частицы~\cite{Saroj2021}, локальный нагрев подложки~\cite{AlMuzaiqer2021}, направление вектора силы тяжести относительно расположения капли и влажность окружающего воздуха~\cite{Du2022,Cedeno2022}. Напряжение электрического тока в подложке влияет на её смачиваемость жидкостью и на режим контактной линии: пиннинг (закрепление границы) или режим постоянного краевого угла (скольжение границы)~\cite{Wang2020}. Эти факторы влияют на геометрию капли, пространственную неоднородность испарения вдоль её свободной поверхности и поле скорости потока жидкости. Такой способ управления можно использовать в режиме реального времени, получая при этом требуемую форму осадка. Еще один дополнительный способ контроля заключается в создании магнитного поля в области капли~\cite{Saroj2021}. В эксперименте~\cite{Saroj2021} в отсутствии магнитного поля формировался кольцевой осадок частиц. Увеличение силы магнитного поля приводило к увеличению количества частиц, осаждающихся в центральной области в виде пятна. Сидячие и висячие капли на подложке изучались в эксперименте при разной влажности воздуха~\cite{Du2022}. Направление вектора силы тяжести относительно расположения капли влияет на то, будут ли поток Марангони и объемная тепловая конвекция сонаправлены или нет. В сидячей капле эти потоки противодействуют друг другу, а в висячей капле наоборот усиливают друг друга. Также не стоит забывать о капиллярном потоке. Влажность воздуха влияет на скорость испарения. Комбинации этих параметров приводили к возникновению разных структур осадков: кольцо, равномерное пятно, диск или центральное пятно~\cite{Du2022}. Коллоидную литографию возможно комбинировать с испарительной литографией~\cite{PerkinsHoward2022}. Для получения масок из микрочастиц на подложке можно использовать испарительную литографию, а затем на их основе с помощью коллоидной литографии получать упорядоченные осадки наночастиц. Эксперимент с сохнущими каплями~\cite{PerkinsHoward2022} показал, что равномерная упорядоченная морфология осадка лучше получается на гидрофильных подложках, чем на гидрофобных. Это связано с отличием поведения контактной линии, что зависит от смачиваемости поверхности.
	
	 Еще одно важное направление заключается в использовании смесей частиц. Эти смеси могут состоять из частиц разного размера\cite{Jose2021,Zolotarev2022}, формы~\cite{Kirner2021}, материала и т.~д. Частицы Януса состоят из двух частей, материалы которых отличаются по физико-химическим свойствам~\cite{Hossain2022}.  Разделение частиц по размеру вблизи трехфазной границы высыхающей капли моделировалось в~\cite{Zolotarev2022}. В бинарных смесях частиц разного размера могут возникать силы исключенного объема (энтропийные силы), которые способствуют притяжению крупных частиц друг к другу~\cite{Mustakim2021,Nozawa2022}. Химическое воздействие на частицы с помощью поверхностно-активных веществ также позволяет контролировать форму осадка и его морфологию~\cite{Galy2022}.
	
	 Испарение коллоидной жидкости из ячейки Хелли-Шоу также можно отнести к испарительной литографии~\cite{Inoue2020a,Homede2020}. Такая ячейка состоит из двух параллельно расположенных пластин, между которыми есть узкая щель, в которой заключен коллоидный раствор. Все боковые стороны ячейки, кроме одной закрыты. Жидкость испаряется через открытое боковое отверстие, поэтому этот процесс называют направленным испарением. Частицы переносятся капиллярным потоком в сторону направленного испарения. Возможно образование как сплошного осадка~\cite{Inoue2020a}, так и периодических полос~\cite{Homede2020}. В~\cite{Homede2020} изучались такие параметры, как степень адсорбции частиц к подложке и скорость испарения, влияющие на морфологию осадка.  Сложная структура возникающих потоков жидкости была показана в эксперименте~\cite{Inoue2020a}.
	
	 На примере капли солевого раствора показана возможность управлять формированием кристаллического осадка с помощью точечного лазерного нагрева локального участка свободной поверхности жидкости~\cite{Li2021a}. В эксперименте изучалось влияние таких параметров как мощность лазера и смачиваемость/ несмачиваемость подложки. Комбинации значений этих параметров приводили к таким кристаллическим паттернам, как кольцо, спираль, пятно и прочее. Визуализация структуры потока показала, что течение направлено в сторону центральной зоны нагрева по направлению от подложки к границе жидкости и воздуха. Вдоль свободной поверхности поток направлен от центра к периферии капли. На перенос растворенного вещества в этом эксперименте в большей степени влиял тепловой поток Марангони и капиллярный поток~\cite{Li2021a}. Во-первых, неравномерный нагрев поверхности приводит к неравномерной локальной плотности потока пара. Во-вторых, перепад температуры влияет на возникновение градиента поверхностного натяжения вдоль свободной поверхности жидкости. Эти два фактора объясняют наблюдаемую структуру потока жидкости. Концентрация соли  растет в области нагрева поверхности не только за счет переноса потоком, но и благодаря интенсивному испарению, происходящему в зоне воздействия лазера~\cite{Li2021a}. Для сравнения в эксперименте~\cite{Shao2021} при испарении капли солевого раствора без какого-либо внешнего воздействия преобладал концентрационный поток Марангони. С помощью метода PIV (Particle Image Velocimetry) была изучена динамика структуры потока. Исследование показало, что зарождение кристаллов и их рост в процессе испарения жидкости приводит к нарушению осесимметричного потока. Вокруг кристаллов возникают симметричные вихри~\cite{Shao2021}.
	
Теоретический интерес к упомянутым выше процессам связан с различными практическими приложениями.  К примеру, в недавней работе~\cite{Hegde2022} проведен эксперимент по управлению формированием осадка, содержащего бактерии, через локальное воздействие на испарение, что важно для приложений в медицине и биотехнологиях. Испарительная литография может использоваться для нанесения функциональных чернил на поверхность и получения необходимого паттерна~\cite{Corletto2021}. Формирование периодических структур из металлических наночастиц важно для разработки плазмонных сенсоров~\cite{Bayat2020}.

\begin{figure}
\includegraphics[width=0.98\textwidth]{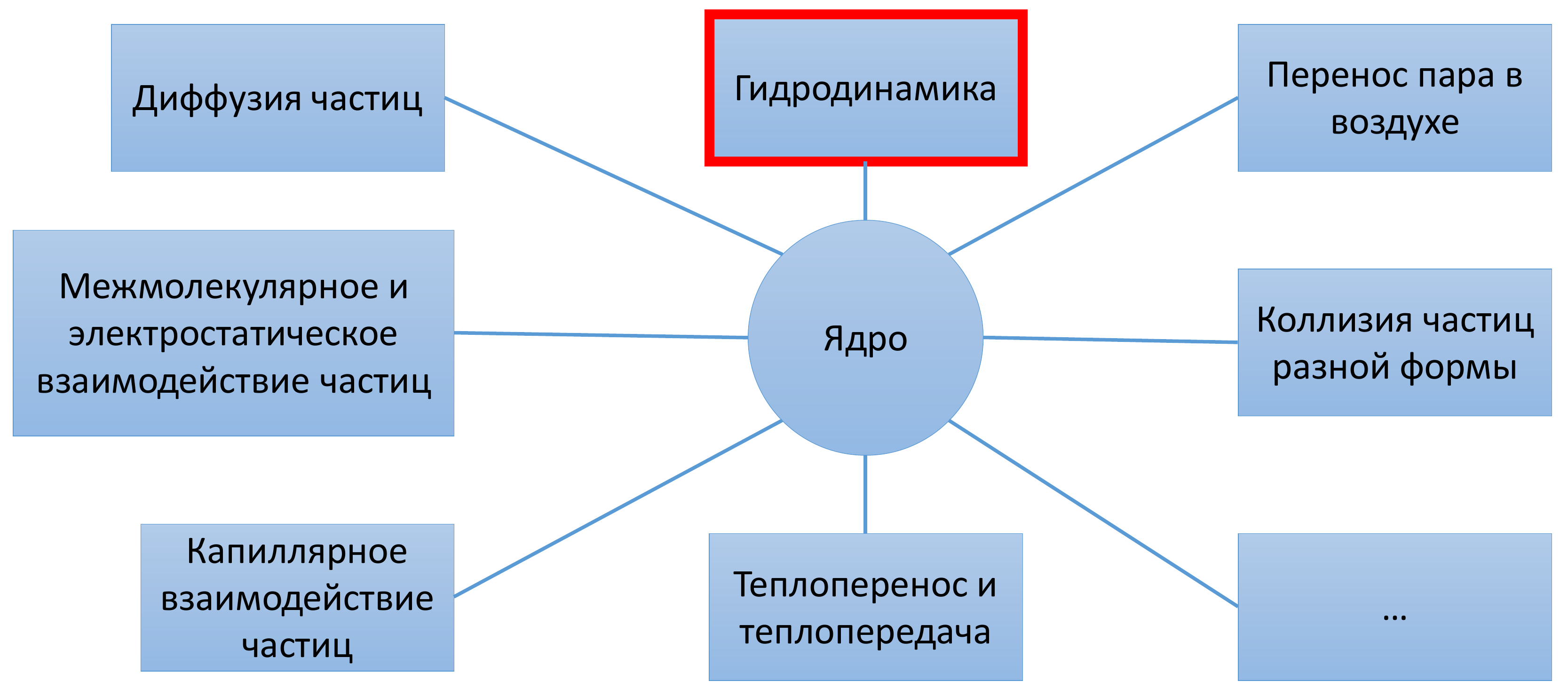}
\caption{Схема планируемого программного комплекса для решения задач в области испарительной литографии}\label{fig:softwarePackage}
\smallskip \footnotesize
[Figure~\ref{fig:softwarePackage}.
Scheme of the planned software package for solving problems in the field of evaporative lithography]
\end{figure}

Моделирование массопереноса в высыхающих каплях важно, так как численные результаты позволяют подобрать необходимые параметры для проведения экспериментальных исследований. Это позволяет совершенствовать существующие методы и разрабатывать новые приложения. Континуальные модели позволяют описать форму осадка, но не его морфологию~\cite{Bhardwaj2010,Tarasevich2011,Kolegov2018344}. Полудискретные модели, в которых частицы описываются точками, тоже не в состоянии предсказывать морфологию осадка~\cite{Petsi2010,Hu2017,Yang2020,Seo2020}. Решеточные модели описывают лишь частицы в форме кубоидов~\cite{Kim2010,Crivoi2014,Zhang2016,Ren2020,Ren2021}. Методы молекулярной динамики и диссипативной динамики частиц позволяют делать прогнозы лишь для относительно малого числа частиц~\cite{LebedevStepanov2013,Breinlinger2014,Liu2019}. Безрешеточные модели на основе метода Монте-Карло, в которых явно отслеживается динамика каждой частицы, лишены упомянутых недостатков~\cite{Andac2019,Kolegov2019JPCS,Lebovka2019,Kolegov2019,Zolotarev2021,Zolotarev2022}. Но эти модели либо не учитывают гидродинамику, либо использую простые аналитические решения для частного случая. Для улучшения этих моделей, алгоритмов и программ необходимо разрабатывать дополнительные модули, позволяющие учесть различные эффекты, влияющие на формирование осадков, в том числе и гидродинамику. В коммерческом пакете Comsol Multiphysics есть модули, позволяющие рассчитывать динамику частиц и гидродинамику~\cite{Song2021}, но цена на этот пакет относительно высокая. Это касается и дополнительных тулбоксов к Matlab~\cite{Marinaro2021}. Недавно для нашей страны был ограничен доступ к официальному сайту свободного пакета для молекулярной динамики LAMMPS (\href{https://www.lammps.org/}{https://www.lammps.org/}). Зарубежные коммерческие пакеты в любой момент могут оказаться недоступными из-за политической ситуации. К примеру, на момент написания данной статьи приостановлены продажи пакета Ansys в РФ. В условиях импортозамещения очень важным является разработка отечественного программного обеспечения для моделирования, написание своих кодов и библиотек. Для проведения исследований в области испарительной литографии требуется разработка программного комплекса, который будет включать ряд модулей (рис.~\ref{fig:softwarePackage}).

Разработка модулей для моделирования диффузии и капиллярного притяжения сферических частиц обсуждалась ранее в~\cite{Kolegov2019JPCS,Kolegov2019,Zolotarev2021}. Также был рассмотрен случай для бинарной смеси частиц разного размера~\cite{Zolotarev2022}. Цель текущей работы~--- разработать численный алгоритм и программный модуль для расчета гидродинамики в высыхающей капле для дальнейшего использования в программном комплексе, ориентированном на решение задач в области испарительной литографии. Но при разработке необходимо учитывать одно требование. Алгоритм должен быть рассчитан на большое количество предельно малых временных шагов (минимальные вычислительные затраты на каждом временном шаге), что позволит корректно работать уже имеющемуся модулю <<Диффузия частиц>>, основанному на методе Монте-Карло, для явного предсказания динамики частиц, так как в этом случае выполняется соотношение Эйнштейна--Смоулховского~\cite{Zolotarev2021}. Величина временного шага $\Delta t$ косвенно зависит от размера частиц, например, для частиц с радиусом 0.35~мкм в~\cite{Zolotarev2021} использовался шаг $\Delta t= 10^{-4}$ с.

Также здесь стоит отметить по какой причине не подходят существующие готовые решения. Описанное в~\cite{Hu2005} аналитическое решение базируется на смеси кинематического подхода и приближения смазки, которое ещё известного как приближение тонкого слоя. Под кинематическим подходом здесь понимается нахождение усреднённой радиальной скорости потока из закона сохранения массы. Такой подход не объясняет природу возникновения потока, как например, капиллярные силы, когда в задаче явным образом учитывают градиент давления Лапласа и кривизну свободной поверхности. Как видно из полученного решения~\cite{Hu2005}, оно явным образом не зависит от ряда физических параметров, как, например, вязкость жидкости. Кроме того, это решение опирается на частный случай с определенным видом функции плотности потока пара $J(r, t)$, что не позволяет решать задачи, связанные с испарительной литографией. Здесь $r$~--- радиальная координата и $t$~--- время. Различные маски, излучатели и прочее внешнее воздействие оказывают влияние на концентрацию пара вблизи свободной поверхности, что в итоге влияет на капиллярный поток жидкости~\cite{Kolegov2020}. Таким образом, для определения $J(r, t)$ необходимо численно моделировать перенос пара в воздухе. Предложенная в~\cite{ParkY2019} неявная разностная схема является неустойчивой. Вблизи границ появляются осцилляции, которые со временем разрастаются на всю область. Авторы~\cite{ParkY2019} подробно не разъясняют способ численной реализации граничных условий, ссылаясь на приложенный код в дополнении. Но восстановить некоторые граничные условия из кода крайне сложно. В программе~\cite{ParkY2019} на каждом временном шаге выполняется искусственное сглаживание осциллирующих точек с помощью специального фильтра. Кроме того, на каждом шаге необходимо выполнять такие вычислительно затратные операции как нахождение обратной матрицы и умножение матрицы на вектор. По этим причинам данный программный модуль~\cite{ParkY2019} нам не подходит. В отличии от~\cite{Tarasevich2011,Kolegov2018344,ParkY2019} здесь дополнительно будет описана численная реализация расчета поля скорости потока жидкости, а не только усредненной по толщине жидкого слоя радиальной компоненты скорости.

\Section{Методы}
\subsection{Физическая постановка задачи}
Капля жидкости испаряется на твёрдом горизонтальном непромокаемом основании при нормальных комнатных условиях. Считаем, что жидкость несжимаема. Трехфазная граница закреплена, поэтому радиус контакта капли и подложки $R$~--- постоянная величина. Рассматриваем случай относительно малого объема жидкости, когда сила поверхностного натяжения преобладает над силой тяжести, число Бонда $\mathrm{Bo} = \rho h_0^2 g/ \sigma \approx 1.4\cdot 10^{-3} \ll 1$, где $g$~--- ускорение свободного падения, $\rho$~--- плотность жидкости, $\sigma$~--- коэффициент поверхностного натяжения и $h_0$~--- начальная высота капли, $h_0= h(r=0,t=0)$. Здесь функция $h(r,t)$ задает профиль свободной поверхности жидкости (граница <<жидкость--воздух>>). В этом случае форма капли близка форме сферического сегмента. По причине осевой симметрии удобно использовать цилиндрическую систему координат, $(r, z)$. Движение жидкости вызывают локальные изменения кривизны поверхности в процессе испарения жидкости, приводящие к возникновению градиента давления Лапласа (капиллярный поток). В случае малых значений краевого угла смачивания $\theta$ капиллярный поток преобладает над потоком Марангони (см. оценку в~\cite{Kolegov2019}). Здесь речь идет о тонких каплях, для которых аспектное отношение $\varepsilon = h_0 / R \ll 1$. Значения параметров задачи приведены в табл.~\ref{tab:parameters}.

\begin{table}[tbp]

\small \centering

\caption{Физические и геометрические параметры задачи [Physical and geometric parameters of the problem] \label{tab:parameters}}
\vspace{-3mm}
\begin{tabular}{p{0.18\textwidth}|p{0.4\textwidth}|p{0.12\textwidth}|p{0.15\textwidth}|}
\hline
\rule{0mm}{11pt}%
\textbf{Обозначение} & \textbf{Расшифровка} & \textbf{Величина} & \textbf{Единицы измерения} \\
\hline
\rule{0mm}{12pt}%
$h_0$ & Высота капли & $10^{-4}$ & м \\
$R$ & Радиус основания капли & $10^{-3}$ & м \\
$h_f$ & Толщина жидкого слоя в районе контактной линии & $0.01 h_0$ & м \\
$\varepsilon= h_0 / R$ & аспектное отношение & 0.1 & -- \\
$\theta \approx 2 \varepsilon$ & Краевой угол смачивания & 0.2 & радианы \\
\hline
$C_0$ & Начальная массовая доля & 0.05 & -- \\
$C_g$ & Концентрация фазового перехода, например, гелеобразования & 0.7 & -- \\
$D$ & Коэффициент диффузии растворенного или взвешенного вещества & $10^{-10}$ & м$^2$/с \\
$\eta_0$ & Вязкость чистой жидкости & $10^{-3}$ & Па$\,$с \\
$\rho$ & Плотность жидкости & $10^3$ & кг/м$^3$ \\
$\sigma$ & Коэффициент поверхностного натяжения & 0.072 & Н/м \\
$t_f$ & Время полного испарения & 450 & с \\
$D_v$ & Коэффициент диффузии пара & $2.4\cdot 10^{-5}$ & м$^2$/с \\
$C_v$ & Плотность насыщенного пара & $2.32\cdot 10^{-2}$ & кг/м$^3$ \\
$H$ & Относительная влажность & 0.4 & -- \\
\hline
\end{tabular}

\end{table}

Распределение растворенного или взвешенного в жидкости вещества будем описывать функцией $\langle C \rangle(r,t)$. Предполагаем, что массовая доля вещества $\langle C \rangle$ не зависит от координаты $z$. Если время диффузионной релаксации вещества много меньше времени полного испарения, $t_d \ll t_f$, то устанавливается равномерная концентрация вещества вдоль вертикального направления. Здесь время $t_d= h_0^2 / D =$ 100~с и время $t_f =$ 450~с. Таким образом, величина $\langle C \rangle$ является усредненной по высоте жидкого слоя.

\subsection{Математическая модель}\label{subsec:mathModel}
Запишем уравнение неразрывности и стационарные уравнения Навье--Стокса без учета инерционных слагаемых в цилиндрической системе координат,
\begin{equation}\label{eq:ContinuityEquation}
  \frac{1}{r} \frac{\partial (r u)}{\partial r} + \frac{\partial w}{\partial z} = 0,
\end{equation}
\begin{equation}\label{eq:NavierStokesEquation_u}
   \frac{\partial}{\partial r}\left(\frac{\eta}{r}\frac{\partial (r u)}{\partial r} \right) + \frac{\partial}{\partial z}\left( \eta \frac{\partial u}{\partial z} \right)= \frac{\partial P}{\partial r},
\end{equation}
\begin{equation}\label{eq:NavierStokesEquation_w}
     \frac{1}{r} \frac{\partial}{\partial r}\left( \eta r \frac{\partial w}{\partial r} \right) + \frac{\partial}{\partial z}\left( \eta \frac{\partial w}{\partial z}\right) = \frac{\partial P}{\partial z},
\end{equation}
где $\eta = \eta(\langle C \rangle)$~--- динамическая вязкость жидкости. Давление $P$, горизонтальная и вертикальная компоненты вектора скорости $\mathbf{v}=(u, w)$ являются функциями, зависящими от $t$, $r$ и $z$.
Система уравнений~\eqref{eq:ContinuityEquation}, \eqref{eq:NavierStokesEquation_u}, \eqref{eq:NavierStokesEquation_w}  справедлива для несжимаемой жидкости и малых значений числа Рейнольдса, $\mathrm{Re}=\rho u_c h_0/ \eta_0$. Как правило, характерная скорость потока жидкости $u_c \approx$ 1~мкм/с в испаряющейся при комнатных условиях капле воды. В таком случае число Рейнольдса $\mathrm{Re} \approx 10^{-4} \ll 1$.

Для перехода к безразмерным записям рассмотрим масштабные соотношения: $\eta = \eta_0 \tilde \eta$, $J = \tilde J J_c$, $u= u_c \tilde u$, $w= \varepsilon u_c \tilde w$, $t= t_c \tilde t$, $P= P_c \, \tilde P$, $r= R \tilde r$ и $z= h_0 \tilde z$. Здесь знаком тильда помечаем безразмерные величины. Заметим, что горизонтальные и вертикальные составляющие (размер, скорость) масштабируются по-разному. Такой подход лежит в основе приближения смазки. Зададим характерные величины: скорость $u_c= \eta_0/ (\rho h_0)\approx$ 0.01~м/с, время $t_c= R/ u_c\approx$ 0.1~с, плотность потока пара $J_c = \varepsilon u_c \rho\approx$ 1~кг/(м$^2$ c) и давление $P_c = R u_c \eta_0/ h_0^2 \approx$ 1~Па. Записываем уравнение неразрывности~\eqref{eq:ContinuityEquation} в безразмерном виде
\begin{equation}\label{eq:ContinuityEquationDimensionless}
  \frac{1}{\tilde r} \frac{\partial (\tilde r \tilde u)}{\partial \tilde r} + \frac{\partial \tilde w}{\partial \tilde z} = 0.
\end{equation}
Заметим, что уравнение~\eqref{eq:ContinuityEquationDimensionless} не отличается по форме записи от~\eqref{eq:ContinuityEquation}.
Уравнения~\eqref{eq:NavierStokesEquation_u} и \eqref{eq:NavierStokesEquation_w} записываются как
\begin{equation}\label{eq:NavierStokesEquation_uDimensionless}
  \varepsilon^2 \frac{\partial}{\partial \tilde r}\left( \frac{\tilde \eta}{\tilde r}\frac{\partial (\tilde r \tilde u)}{\partial \tilde r} \right) + \frac{\partial}{\partial \tilde z}\left( \tilde \eta \frac{\partial \tilde u}{\partial \tilde z}\right) = \frac{\partial \tilde P}{\partial \tilde r},
\end{equation}
\begin{equation}\label{eq:NavierStokesEquation_wDimensionless}
   \varepsilon^3 \frac{1}{\tilde r} \frac{\partial}{\partial \tilde r}\left( \tilde \eta \tilde r \frac{\partial \tilde w}{\partial \tilde r} \right) + \varepsilon^2 \frac{\partial}{\partial \tilde z} \left( \tilde \eta \frac{\partial \tilde w}{\partial \tilde z}\right) = \frac{\partial \tilde P}{\partial \tilde z}.
\end{equation}
Пренебрегаем слагаемыми, при которых стоят $\varepsilon^2$ и $\varepsilon^3$, в~\eqref{eq:NavierStokesEquation_uDimensionless} и \eqref{eq:NavierStokesEquation_wDimensionless}, получаем
\begin{equation}\label{eq:NavierStokesEquation_uDimensionless2}
   \frac{\partial \tilde P}{\partial \tilde r} - \frac{\partial}{\partial \tilde z}\left( \tilde \eta \frac{\partial \tilde u}{\partial \tilde z}\right) = 0,
\end{equation}
\begin{equation}\label{eq:NavierStokesEquation_wDimensionless2}
  \frac{\partial \tilde P}{\partial \tilde z} = 0.
\end{equation}
В результате получили упрощенную систему уравнений~\eqref{eq:ContinuityEquationDimensionless}, \eqref{eq:NavierStokesEquation_uDimensionless2} и \eqref{eq:NavierStokesEquation_wDimensionless2}.

Из~\eqref{eq:NavierStokesEquation_wDimensionless2} следует, что давление $\tilde P$ не зависит от $\tilde z$, тогда вместо $\tilde P$ будем рассматривать усредненную по высоте жидкого слоя величину $\langle \tilde P \rangle = \langle \tilde P \rangle (\tilde r, \tilde t)$.  Проинтегрируем левую и правую части~\eqref{eq:NavierStokesEquation_uDimensionless2},
учитывая, что вязкость $\tilde \eta$ здесь не зависит явно или косвенно от $\tilde z$,
$$\frac{\partial \langle \tilde P \rangle}{\partial \tilde r}\int d\tilde z - \tilde \eta \int \frac{\partial^2 \tilde u}{\partial \tilde z^2}\,d\tilde z = \int 0\,d\tilde z, $$
получаем
$$\frac{\partial \langle \tilde P \rangle}{\partial \tilde r} (\tilde z + C_1) - \tilde \eta \left( \frac{\partial \tilde u}{\partial \tilde z} + C_2 \right) = C_3.$$
Как было сказано ранее, термокапиллярные потоки здесь не рассматриваются, поэтому на свободной границе, $\tilde z = \tilde h$, выполняется баланс касательных напряжений $\partial \tilde u/ \partial \tilde z =0$, тогда
$$\frac{\partial \langle \tilde P \rangle}{\partial \tilde r} \tilde h + \frac{\partial \langle \tilde P \rangle}{\partial \tilde r} C_1  = C_3 + \tilde \eta C_2.$$
Для $\tilde r = 0$ выполняется условие осевой симметрии $\partial \langle \tilde P \rangle/ \partial \tilde r = 0$, тогда $C_3 = - \tilde \eta(\tilde r=0, \tilde t) C_2$. В действительности вязкость $\tilde \eta$ зависит от $\tilde r$ и $\tilde t$ неявным образом, через массовую долю $\langle C \rangle$, но здесь для краткости пишем $\tilde \eta(\tilde r, \tilde t)$. На границе $\tilde r = \tilde R$ горизонтальная компонента скорости является постоянной величиной, не зависящей от $\tilde z$ (в силу прилипания к подложке $\tilde u = 0$). Тогда в этой точке $\partial^2 \tilde u/ \partial \tilde z^2 = 0$. Учитывая это, из~\eqref{eq:NavierStokesEquation_uDimensionless2} получаем граничное условие $\partial \langle \tilde P \rangle / \partial \tilde r = 0$ для $\tilde r = \tilde R$. С таким условием приходим к соотношению $C_3 = - \tilde \eta(\tilde r=\tilde R, \tilde t) C_2$. На разных границах получили одно и тоже соотношение, хотя в общем случае $\tilde \eta(\tilde r=0, \tilde t) \neq \tilde \eta(\tilde r = \tilde R, \tilde t)$. Приходим к выводу, что $C_2 = C_3 = 0$. В таком случае константа интегрирования $C_1 = - \tilde h$, тогда выражение принимает вид
$$\frac{\partial \langle \tilde P \rangle}{\partial \tilde r} (\tilde z - \tilde h) - \tilde \eta \frac{\partial \tilde u}{\partial \tilde z} = 0.$$
Интегрируем еще раз,
$$\frac{\partial \langle \tilde P \rangle}{\partial \tilde r} \int \tilde z\,d\tilde z - \frac{\partial \langle \tilde P \rangle}{\partial \tilde r} \tilde h \int d\tilde z - \tilde \eta \int \frac{\partial \tilde u}{\partial \tilde z}\,d\tilde z = \int 0\,d\tilde z,$$
получаем
$$\frac{\partial \langle \tilde P \rangle}{\partial \tilde r}\left( \frac{\tilde z^2}{2} + C_4 - \tilde h \tilde z +C_5 \right) - \tilde \eta (\tilde u + C_6) = C_7.$$ Учитывая, что для $\tilde z = 0$ выполняется условие прилипания $\tilde u = 0$ и для $\tilde r = 0$ выполняется условие $\partial \langle \tilde P \rangle/ \partial \tilde r = 0$, получаем $C_7 = - \tilde \eta(\tilde r = 0, \tilde t) C_6$. Учитывая аналогичные граничные условия для $\tilde r = \tilde R$, получаем $C_7 = - \tilde \eta(\tilde r = \tilde R, \tilde t) C_6$. Приходим к выводу, что $C_6 = C_7 = 0$.  Затем, учитывая лишь условие $\tilde u = 0$ для $\tilde z = 0$, получаем $C_4 = - C_5$. В итоге выражаем горизонтальную компоненту вектора скорости,
\begin{equation}\label{eq:uVelocityDimensionless}
   \tilde u = \frac{H_a(x)}{\tilde \eta} \frac{\partial \langle \tilde P \rangle}{\partial \tilde r}\left( \frac{\tilde z^2}{2} - \tilde h \tilde z \right).
\end{equation}
В некоторых уравнениях модели, в том числе и в~\eqref{eq:uVelocityDimensionless}, дополнительно используется аналитическое приближение функции Хевисайда $H_a$, чтобы исключить некоторые виды массопереноса при возникновении фазового перехода, как, например, <<золь--гель>>, <<жидкость--стекло>> и т.~п., $$H_a(x) = \frac{1}{1+\exp(-2 k_h x)},$$ где $x = C_g - \langle C \rangle - \delta x$, $k_h = 10/\delta x$ ($\delta x$~--- ширина области перехода).
Теперь подставляем~\eqref{eq:uVelocityDimensionless} в \eqref{eq:ContinuityEquationDimensionless} и интегрируем,
$$\int \frac{1}{\tilde r} \frac{\partial}{\partial \tilde r} \left( \frac{\tilde r H_a(x)}{\tilde \eta}  \frac{\partial \langle \tilde P \rangle}{\partial \tilde r} \left( \frac{\tilde z^2}{2} - \tilde h \tilde z \right) \right)\,d \tilde z + \int  \frac{\partial \tilde w}{\partial \tilde z}\,d \tilde z = \int 0\,d \tilde z,$$
получаем
$$\frac{1}{\tilde r} \frac{\partial}{\partial \tilde r} \left( \frac{\tilde r H_a(x)}{\tilde \eta}  \frac{\partial \langle \tilde P \rangle}{\partial \tilde r} \left( \frac{\tilde z^3}{6} - \frac{\tilde h \tilde z^2}{2} + C_8 \right) \right) + \tilde w + C_9 = 0.$$
Запишем полученное выражение в развернутом виде,
\begin{multline*}
  \frac{H_a(x)}{\tilde r \tilde \eta} \frac{\partial \langle \tilde P \rangle}{\partial \tilde r} \left( \frac{\tilde z^3}{6} - \frac{\tilde h \tilde z^2}{2} \right) + \frac{H_a(x)}{\tilde \eta} \left( \frac{\tilde z^3}{6} - \frac{\tilde h \tilde z^2}{2} \right)  \frac{\partial^2 \langle \tilde P \rangle}{\partial \tilde r^2} - \frac{H_a(x) \tilde z^2}{2 \tilde \eta} \frac{\partial \langle \tilde P \rangle}{\partial \tilde r} \frac{\partial \tilde h}{\partial \tilde r} + \\
  + \frac{C_8 H_a(x)}{\tilde r \tilde \eta} \frac{\partial \langle \tilde P \rangle}{\partial \tilde r} + \frac{C_8 H_a(x)}{\tilde \eta} \frac{\partial^2 \langle \tilde P \rangle}{\partial \tilde r^2} + \\ + \frac{\partial H_a(x)}{\partial \tilde r} \frac{1}{\tilde \eta}  \frac{\partial \langle \tilde P \rangle}{\partial \tilde r} \left( \frac{\tilde z^3}{6} - \frac{\tilde h \tilde z^2}{2} + C_8 \right) + \tilde w + C_9 = 0.
\end{multline*}
Учитывая условие непротекания, $\tilde w = 0$ для $\tilde z = 0$, и условие $\partial \langle \tilde P \rangle/ \partial \tilde r = 0$ для $\tilde r = 0$ и $\tilde r = \tilde R$, получаем два соотношения:
$$\frac{C_8 H_a(x)}{\tilde \eta(\tilde r = 0, \tilde t)} \frac{\partial^2 \langle \tilde P \rangle(\tilde r = 0, \tilde t)}{\partial \tilde r^2} + C_9 = 0 \text{ и } \frac{C_8 H_a(x)}{\tilde \eta(\tilde r = \tilde R, \tilde t)} \frac{\partial^2 \langle \tilde P \rangle(\tilde r = \tilde R, \tilde t)}{\partial \tilde r^2} + C_9 = 0.$$
Из чего следует, что $C_8 = C_9 =0$. В таком случае получаем выражение для вертикальной компоненты вектора скорости,
\begin{equation}\label{eq:wVelocityDimensionless}
   \tilde w = -\frac{1}{\tilde r} \frac{\partial}{\partial \tilde r}\left( \frac{\tilde r H_a(x)}{\tilde \eta} \frac{\partial \langle \tilde P \rangle}{\partial \tilde r} \left( \frac{\tilde z^3}{6} - \frac{\tilde h \tilde z^2}{2} \right) \right).
\end{equation}

Движение свободной поверхности капли в процессе испарения описывается законом сохранения (см. ссылки в~\cite{Kolegov2018344})
\begin{equation}\label{eq:DropletThicknessEvolution}
   \frac{\partial h}{\partial t} + \frac{1}{r} \frac{\partial (rh\left\langle u\right\rangle)}{\partial r} = -\frac{J}{\rho} \sqrt{1+\left(\frac{\partial h}{\partial r}\right)^2},
\end{equation}
где $\left\langle u\right\rangle$~--- усредненная по высоте жидкого слоя скорость радиального потока жидкости. В безразмерном виде~\eqref{eq:DropletThicknessEvolution} записывается как
\begin{equation}\label{eq:DropletThicknessEvolutionDimensionless}
   \frac{\partial \tilde h}{\partial \tilde t} + \frac{1}{\tilde r} \frac{\partial (\tilde r \tilde h\left\langle \tilde u\right\rangle)}{\partial \tilde r} = - \tilde J,
\end{equation}
где пренебрегаем $\sqrt{1+\varepsilon^2 \left(\partial \tilde h/ \partial \tilde r\right)^2}$ в правой части как величиной второго порядка малости. С учетом~\eqref{eq:uVelocityDimensionless} усредненную радиальную скорость $\left\langle \tilde u\right\rangle$ выражаем как
\begin{equation}\label{eq:AveragedRadialVelocityDimensionless}
\left\langle \tilde u\right\rangle = \frac{1}{\tilde h} \int \limits_{\tilde z=0}^{\tilde z = \tilde h} \tilde u\, d\tilde z = -H_a(x)\frac{\tilde h^2}{3 \tilde \eta} \frac{\partial \langle \tilde P \rangle}{\partial \tilde r}.
\end{equation}
Капиллярное давление зависит от локальной кривизны поверхности,
\begin{equation}\label{eq:CapillaryPressureDimensionless}
  \langle \tilde P \rangle = - \frac{1}{\mathrm{Ca}} \frac{1}{\tilde r} \frac{\partial}{\partial \tilde r} \left( \tilde r \frac{\partial \tilde h}{\partial \tilde r} \right),
\end{equation}
что следует из формулы для давления Лапласа~\cite{Kolegov2018344} с учетом приближения смазки. Здесь капиллярное число $\mathrm{Ca}= \eta_0 u_c/ (\sigma \varepsilon^3) = \eta^2 /(\varepsilon^3 \rho h_0 \sigma) \approx$ 0.14, где $\sigma$~--- коэффициент поверхностного натяжения. В~\cite{ParkY2019} выражения~\eqref{eq:AveragedRadialVelocityDimensionless} и \eqref{eq:CapillaryPressureDimensionless} подставляются в уравнение~\eqref{eq:DropletThicknessEvolutionDimensionless}, и затем строится разностная схема для уравнения с производной четвертого порядка по $r$ относительно $h$. Чтобы избежать таких громоздких разностных схем, здесь~\eqref{eq:AveragedRadialVelocityDimensionless} и \eqref{eq:CapillaryPressureDimensionless} рассматриваются как отдельные уравнения.

Для описания переноса растворённого или взвешенного вещества используем уравнение конвекции--диффузии~\cite{Kolegov2018344}
\begin{equation}\label{eq:ConvectionDiffusionEquation}
  \frac{\partial \left\langle C \right\rangle}{\partial t} + \left\langle u\right\rangle \frac{\partial \left \langle C \right\rangle}{\partial r} = \frac{D}{r h} \frac{\partial}{\partial r} \left( r h \frac{\partial \left\langle C \right\rangle}{\partial r} \right) + \frac{J \left \langle C \right\rangle}{\rho h} \sqrt{1+\left(\frac{\partial h}{\partial r}\right)^2},
\end{equation}
где $D$~--- коэффициент диффузии растворенного или взвешенного вещества. Подробный вывод уравнения~\eqref{eq:ConvectionDiffusionEquation} описан в~\cite{KolegovThesis2018}. В безразмерном виде уравнение~\eqref{eq:ConvectionDiffusionEquation} с учетом приближения смазки принимает вид
\begin{equation}\label{eq:ConvectionDiffusionEquationDimensionless}
  \frac{\partial \left\langle C \right\rangle}{\partial \tilde t} + \left\langle \tilde u\right\rangle \frac{\partial \left \langle C \right\rangle}{\partial \tilde r} = \frac{H_a(x)}{\mathrm{Pe}} \frac{1}{\tilde r \tilde h} \frac{\partial}{\partial \tilde r} \left( \tilde r \tilde h \frac{\partial \left\langle C \right\rangle}{\partial \tilde r} \right) + \frac{\tilde J \left \langle C \right\rangle}{\tilde h},
\end{equation}
где $\mathrm{Pe}$~--- число Пекле, $\mathrm{Pe}\approx R u_c/ D\approx 10^5$. В итоге система уравнений включает уравнения~\eqref{eq:uVelocityDimensionless}, \eqref{eq:wVelocityDimensionless}, \eqref{eq:DropletThicknessEvolutionDimensionless}, \eqref{eq:AveragedRadialVelocityDimensionless}, \eqref{eq:CapillaryPressureDimensionless} и \eqref{eq:ConvectionDiffusionEquationDimensionless}.

\subsection{Вязкость и плотность потока пара}
Теперь необходимо ввести замыкающие соотношения для некоторых величин. В предложенной здесь модели вязкость зависит от массовой доли вещества. Для описания этой зависимости используем формулу Муни
$$\tilde \eta = \exp \left(\frac{S \left\langle C \right\rangle}{1 - K \left\langle C \right\rangle}\right),$$
где значения эмпирических параметров $S$ и $K$ возьмем из~\cite{Kolegov2018113}, $S = 1.692$ и $K = 1.236$.

Также считаем, что плотность потока пара зависит от концентрации вещества и толщины жидкого слоя,
$$\tilde J = \tilde J_0 \frac{ 1 - \left\langle C \right\rangle^2 / C_g^2}{\kappa + \tilde h},$$
где $C_g$~--- критическая массовая доля, при
которой происходит фазовый переход~\cite{Tarasevich2011} (золь--гель, стеклообразование, кристаллизация и т.~п.), предположим $C_g \approx 0.7$~\cite{Kolegov2018113}. Для расчетов будем использовать значение параметра $\kappa = 1$. Это некоторая аппроксимация для учета испарения, поэтому в дальнейшем планируется разработать модуль численного расчета переноса пара в воздухе (рис.~\ref{fig:softwarePackage}). Величина $\tilde J_0$ рассчитывается как
$$
\tilde J_0 = \frac{D_v C_v (1 - H)}{J_c R} (0.27 \theta^2 + 1.3)(0.6381 - 0.2239(\theta - \pi / 4)^2),
$$
где $H$~--- относительная влажность, $D_v$~--- коэффициент диффузии пара и $C_v$~--- плотность насыщенного пара~\cite{Hu2005}.

\subsection{Начальные и граничные условия}\label{subsec:IBC}
Запишем начальные условия задачи.
Для тонких капель, размер которых меньше капиллярной длины, допустимо использовать параболическое приближение начальной формы свободной поверхности $\tilde h(\tilde r, \tilde t=0)= 1 - \tilde r^2$~\cite{Hu2005}.
Начальную массовую долю вещества опишем с помощью выражения~\cite{Tarasevich2011}
$$\left\langle C \right\rangle (\tilde r, \tilde t=0) = C_g \frac{2 - C_0 + 2(C_0 -1)}{1 + \exp(w_c (\tilde r - 1))},$$
где параметр $w_c$ регулирует начальную ширину кольцевого осадка (здесь в расчетах используем $w_c = 30$). Такое условие предполагает, что вещество в начале процесса распределено практически равномерно вдоль радиальной координаты за исключением района периферии ($\tilde r \approx \tilde R = 1$), где концентрация устремляется к пороговому значению $C_g$, что обеспечивает закрепление трехфазной границы <<жидкость--подложка--воздух>>.

Для $\tilde r=0$ по причине осевой симметрии задаются следующие граничные условия:
$$\frac{\partial \langle \tilde P \rangle}{\partial \tilde r} = \frac{\partial \tilde h}{\partial \tilde r} = \frac{\partial \langle C \rangle}{\partial \tilde r}= \frac{\partial \tilde w}{\partial \tilde r} = \tilde u = \langle \tilde u \rangle = 0.$$

На границе $\tilde r= \tilde R$ выполняются следующие условия:
$$\frac{\partial \langle \tilde P \rangle}{\partial \tilde r} = \tilde w = \tilde u = \langle \tilde u \rangle = 0, \tilde h = \tilde h_f,  \langle C \rangle = C_g.$$ Условие для $\langle \tilde P \rangle$ уже обсуждалось в~\S~\ref{subsec:mathModel}. Кроме того, здесь учитываем отсутствие потока растворенного или взвешенного вещества через границу $\tilde r= \tilde R$. Вертикальная и горизонтальная компоненты вектора скорости потока жидкости записываются из соображений непротекания и прилипания. Здесь $\tilde h_f$~--- толщина жидкого слоя в районе контактной линии.

Граничные условия вдоль подложки следующие ($\tilde z = 0$),
$$\tilde w = \tilde u = 0,$$
по причине не протекания и прилипания. Вдоль свободной границы жидкости условия принимают вид ($\tilde z = \tilde h$)
$$\frac{\partial \tilde u}{\partial \tilde z} = \tilde w = 0,$$
считая, что на этой границе нет вязкого трения и отсутствует поток жидкости через границу <<жидкость--воздух>> (из-за медленного испарения влияние подвижной границы и отдачи пара не учитываем). В приближении смазки на этой границе $\tilde w$ и $\tilde u$ соответствуют нормальной и касательной компонентам вектора скорости. Последние условия для скорости потока на свободной поверхности капли являются излишними, так как значения $\tilde u$ и $\tilde w$ на этой границе не сложно рассчитать с использованием выражений~\eqref{eq:uVelocityDimensionless} и \eqref{eq:wVelocityDimensionless}.

\subsection{Численный метод решения задачи}
В этом разделе работаем с безразмерными величинами, поэтому знак <<тильда>> опускаем для краткости. Для дискретной аппроксимации уравнений модели с помощью разностных схем введем  пространственно-временную сетку с координатами $r_n = n\, \Delta r$, $z_m = m\, \Delta z$ и $t_k = k\, \Delta t$, где $n= 0, 1, \dots , N$; $m= 0, 1, \dots , M$; $k= 0, 1, \dots , K$. Параметры $\Delta t$, $\Delta r$ и $\Delta z$ обозначают шаги по времени и пространству, соответственно. Кроме того, сетка характеризуется такими величинами, как количество отрезков ($N$, $M$ и $K$) и узлов ($N+1$, $M+1$ и $K+1$). Параметр $K = [t_f / \Delta t]$, здесь квадратные скобки обозначают целочисленное деление. Шаги по пространству определяются как $\Delta r= R/N$ и $\Delta z= h_0/M$.

Перепишем уравнения~\eqref{eq:uVelocityDimensionless}, \eqref{eq:wVelocityDimensionless}, \eqref{eq:DropletThicknessEvolutionDimensionless}, \eqref{eq:AveragedRadialVelocityDimensionless},  \eqref{eq:CapillaryPressureDimensionless} и \eqref{eq:ConvectionDiffusionEquationDimensionless} в дискретном виде:
\begin{equation}\label{eq:uVelocityDiscrete}
   u_{n,m}^k = \frac{H_a(x_n)}{\eta_n^k} \frac{\langle P \rangle_{n+1}^k - \langle P \rangle_{n-1}^k}{2\, \Delta r}\left( \frac{z_m^2}{2} - h_n^k z_m \right),
\end{equation}
\begin{multline}\label{eq:wVelocityDiscrete}
    w_{n,m}^k = -\frac{H_a(x_{n+0.5})}{r_n\, \Delta r}  \frac{r_{n+0.5}}{\eta_{n+0.5}^k} \frac{ \langle P \rangle_{n+1}^k - \langle P \rangle_n^k}{\Delta r} \left( \frac{z_m^3}{6} - \frac{h_{n+0.5}^k z_m^2}{2} \right) +\\ +\frac{H_a(x_{n-0.5})}{r_n\, \Delta r} \frac{r_{n-0.5}}{\eta_{n-0.5}^k} \frac{ \langle P \rangle_n^k - \langle P \rangle_{n-1}^k}{\Delta r} \left( \frac{z_m^3}{6} - \frac{h_{n-0.5}^k z_m^2}{2} \right),
\end{multline}
\begin{equation}\label{eq:DropletThicknessEvolutionDiscrete}
   \frac{h_n^{k+1} - h_n^k}{\Delta t} + \frac{1}{r_n} \frac{r_{n+1}  h_{n+1}^{k+1} \langle u\rangle_{n+1}^{k+1} - r_{n-1}  h_{n-1}^{k+1} \langle u\rangle_{n-1}^{k+1}}{2\Delta r} = - J_n^k,
\end{equation}
\begin{equation}\label{eq:AveragedRadialVelocityDiscrete}
\langle u\rangle_n^k = -H_a(x_n)\frac{(h_n^k)^2}{3 \eta_n^k} \frac{\langle P \rangle_{n+1}^k - \langle P \rangle_{n-1}^k}{2 \Delta r},
\end{equation}
\begin{multline}\label{eq:ConvectionDiffusionEquationDiscrete}
  \frac{\langle C \rangle_n^{k+1} - \langle C \rangle_n^k}{\Delta t} + \frac{|\langle u\rangle_n^{k+1}|+\langle u\rangle_n^{k+1}}{2} \frac{ \langle C \rangle_n^k - \langle C \rangle_{n-1}^k}{\Delta r} - \\- \frac{|\langle u\rangle_n^{k+1}|-\langle u\rangle_n^{k+1}}{2} \frac{ \langle C \rangle_{n+1}^k - \langle C \rangle_n^k}{\Delta r} = \frac{J_n^k \langle C \rangle_n^k}{h_n^{k+1}} + \\ \frac{1}{\mathrm{Pe}} \frac{H_a(x_n)}{r_n h_n^{k+1} \Delta r} \left( r_{n+0.5} h_{n+0.5}^{k+1} \frac{ \langle C \rangle_{n+1}^k - \langle C \rangle_n^k }{\Delta r} - r_{n-0.5} h_{n-0.5}^{k+1} \frac{ \langle C \rangle_n^k - \langle C \rangle_{n-1}^k }{\Delta r}\right),
\end{multline}
\begin{equation}\label{eq:CapillaryPressureDiscrete}
  \langle P \rangle_n^k = -\frac{1}{\mathrm{Ca}} \frac{1}{\Delta r\, r_n} \left( r_{n+0.5} \frac{h_{n+1}^k - h_n^k}{\Delta r} -r_{n-0.5} \frac{h_n^k - h_{n-1}^k}{\Delta r} \right),
\end{equation}
где $x_n = C_g - \langle C \rangle_n^k - \delta x$ и $k_h = 10/\delta x = 2000$. Промежуточные значения между узлами вида $f_{n\pm 0.5}$ рассчитываются как $(f_{n\pm 1} + f_n)/2$. Стоит заметить, что в~\eqref{eq:ConvectionDiffusionEquationDiscrete} для дискретизации конвективного слагаемого используется схема с разностями против потока. Для приближения пространственной производной первого порядка в~\eqref{eq:uVelocityDiscrete}, \eqref{eq:DropletThicknessEvolutionDiscrete} и \eqref{eq:AveragedRadialVelocityDiscrete} используется центральная разностная схема. Это даёт второй порядок аппроксимации по $\Delta r$, как и аппрокисмация производной второго порядка в~\eqref{eq:wVelocityDiscrete}, \eqref{eq:ConvectionDiffusionEquationDiscrete} и \eqref{eq:CapillaryPressureDiscrete}.
Здесь используется как явная разностная схема~\eqref{eq:ConvectionDiffusionEquationDiscrete}, так и неявная~---  \eqref{eq:DropletThicknessEvolutionDiscrete}.

Теперь запишем начальные и граничные условия из раздела~\ref{subsec:IBC} в дискретном виде. Начальная форма свободной поверхности в дискретном виде задается как $h_n^0 = 1 - r_n^2$. Формула начального распределения массовой доли частиц принимает вид
$$\langle C \rangle_n^0 = C_g \frac{2 - C_0 + 2(C_0 -1)}{1 + \exp(w_c (r_n - 1))}.$$

С граничными условиями первого рода всё просто, $f_{n,m}^k = c$, где $f$~---искомая функция и $c$~--- некоторая константа (здесь $n=0$ или $n=N$). Для граничных условий второго рода, вида $\partial f/ \partial r = 0$, здель используются аппроксимации второго порядка точности: $f_{0,m} = (4 f_{1,m} - f_{2,m}) / 3$, $f_{N,m} = 4 f_{N-1,m} - 3 f_{N-2,m}$. Для величин, не зависящих от $z$, индекс $m$ в этих формулах отсутствует.

Дискретное уравнение~\eqref{eq:DropletThicknessEvolutionDiscrete} можно расщепить по физическим процессам на два уравнения. Одно из них описывает изменение $h$ в результате испарения,
\begin{equation}\label{eq:DropletThicknessEvolutionEvaporationDiscrete}
	\frac{h_n^\mathrm{tmp} - h_n^k}{\Delta t} = - J_n^k,
\end{equation}
другое~--- в результате конвективного переноса массы,
\begin{equation}\label{eq:DropletThicknessEvolutionConvectionDiscrete}
	\frac{h_n^{k+1} - h_n^\mathrm{tmp}}{\Delta t} + \frac{1}{r_n} \frac{r_{n+1}  h_{n+1}^{k+1} \langle u\rangle_{n+1}^{k+1} - r_{n-1}  h_{n-1}^{k+1} \langle u\rangle_{n-1}^{k+1}}{2\Delta r} = 0.
\end{equation}
Здесь~\eqref{eq:DropletThicknessEvolutionEvaporationDiscrete} и \eqref{eq:DropletThicknessEvolutionConvectionDiscrete}~--- явная и неявная разностные схемы. Временное значение высоты капли $h_n^\mathrm{tmp}$ определяется из~\eqref{eq:DropletThicknessEvolutionEvaporationDiscrete}. Затем, после пересчёта давления~\eqref{eq:CapillaryPressureDiscrete} и усредненной скорости~\eqref{eq:AveragedRadialVelocityDiscrete}, решается~\eqref{eq:DropletThicknessEvolutionConvectionDiscrete} и находится итоговое значение $h_n^{k+1}$ на новом временном слое. Уравнение~\eqref{eq:DropletThicknessEvolutionConvectionDiscrete} можно переписать в виде
\begin{equation}\label{eq:DropletThicknessEvolutionConvectionTridiagonalDiscrete}
a_n h_{n-1}^{k+1} + b_n h_n^{k+1} + c_n h_{n+1}^{k+1} = d_n,
\end{equation}
где $a_n = - \Delta t \, r_{n-1} \langle u\rangle_{n-1}^{k+1}/(2 r_n \, \Delta r)$, $b_n = 1$, $c_n = \Delta t \, r_{n+1} \langle u\rangle_{n+1}^{k+1}/(2 r_n \, \Delta r)$ и $d_n = h_n^\mathrm{tmp}$. В совокупности для всех узлов $n$ из~\eqref{eq:DropletThicknessEvolutionConvectionTridiagonalDiscrete} получаем СЛАУ, которую можно представить в матричной форме,
$$
\begin{pmatrix}
	b_0& c_0 & 0 & 0 &\ldots & 0\\
	a_1& b_1 & c_1 & 0 &\ldots & 0\\
	0 & a_2 & b_2 & c_2 &\ldots & 0\\
	\vdots& \vdots& \ddots &\ddots &\ddots & \vdots\\
	0 & \ldots & 0 & a_{N-1} & b_{N-1} & c_{N-1}\\
	0 & 0 & 0 & \ldots & a_N & b_N
\end{pmatrix}
\cdot
\begin{pmatrix}
	h_0^{k+1}\smallskip\\
	h_1^{k+1}\\
	\vdots\\
	h_{N-1}^{k+1}\smallskip\\
	h_N^{k+1}
\end{pmatrix}
=
\begin{pmatrix}
	d_0\smallskip\\
	d_1\smallskip\\
	\vdots\smallskip\\
	d_{N-1}\smallskip\\
	d_N
\end{pmatrix}
.
$$
Запишем в более компактном виде, $\mathbf{A}\cdot \mathbf{h} = \mathbf{d}$, где $\mathbf{A}$~--- матрица коэффициентов, $\mathbf{d}$~--- вектор правых частей и $\mathbf{h}$~--- искомый вектор. Обратим внимание, что коэффициенты $a_0$ и $c_N$ в вычислениях не задействовываются, и квадратная матрица $\mathbf{A}$ характеризуется ленточной трехдиагональной структурой ненулевых элементов, поэтому такая СЛАУ может быть эффективно решена методом прогонки. На основании граничных условий запишем значения коэффицентов $b_0$, $c_0$, $d_0$, $a_N$, $b_N$ и $d_N$. Чтобы не нарушить трехдиагональную структуру матрицы $\mathbf{A}$, для граничного условия на оси симметрии $\partial h/ \partial r = 0$ используем аппроксимацию первого порядка, $(h_1^{k+1} - h_0^{k+1})/ \Delta t \approx 0$. В таком случае получаем коэффициенты $b_0 = 1$, $c_0 = -1$ и $d_0 = 0$. Условие на трехфазной границе $h = h_f$ приводит к коэффициентам $a_N = 0$, $b_N = 1$ и $d_N = h_f$.

Значения $h_n^{k+1}$, $\langle u\rangle_{n}^{k+1}$ и $\langle P\rangle_{n}^{k+1}$ необходимо итеративно уточнять. На первой итерации задаем отправное значение $h_n^{k+1,i=1} = h_n^\mathrm{tmp}$. Относительно резкое изменение давления за дискретный временной шаг в результате испарения согласно уравнению~\eqref{eq:DropletThicknessEvolutionEvaporationDiscrete} приводит к неустойчивости вычислений. По этой причине здесь используется метод явной релаксации, $$\langle P\rangle_{n}^{k+1,i} = \langle P\rangle_{n}^{k+1,i-1} + \alpha \left(\langle P\rangle_{n}^{k+1,i} - \langle P\rangle_{n}^{k+1,i-1}\right),$$ где $i$~--- номер итерации и $\alpha$~--- фактор релаксации. Предварительно $\langle P\rangle_{n}^{k+1,i}$ рассчитывается по формуле~\eqref{eq:CapillaryPressureDiscrete}. Значение фактора релаксации подобрано эмпирическим путем. Здесь в расчетах использовалось значение $\alpha= 10^{-3}$. Малое значение этого параметра улучшает устойчивость численного счёта, но при этом увеличивает количество итераций, необходимых для сходимости решения. Вблизи границ при расчёте усреднённой скорости $\langle u\rangle$ по формуле~\eqref{eq:AveragedRadialVelocityDiscrete} возникают осцилляции, которые за несколько временных шагов приводят к <<разрушению>> вычислений. С целью предотвращения таких осцилляций для приграничных узлов использовалась линейная интерполяция. Учитывая, что $\langle u\rangle \to 0$ при $r \to 0$ или при $r \to R$, получаем $\langle u\rangle_1 = \langle u\rangle_2 / 2$ и $\langle u\rangle_{N-1} = \langle u\rangle_{N-2} / 2$.

Механизм остановки итераций включает два возможных события: достигнуто максимальное число итераций $I_\mathrm{max}=100$ или максимальная невязка стала меньше заданной погрешности, $r_\mathrm{max} < \epsilon_\mathrm{lim}$, где $\epsilon_\mathrm{lim} = 10^{-17}$ и
$$r_\mathrm{max} = \left| \underset{n}{\max} \left( h_n^{k+1} - h_n^\mathrm{tmp} + \frac{\Delta t}{r_n} \frac{r_{n+1} h_{n+1}^{k+1} \langle u\rangle_{n+1}^{k+1} - r_{n-1}  h_{n-1}^{k+1} \langle u\rangle_{n-1}^{k+1}}{2\Delta r} \right)\right|$$ согласно уравнению~\eqref{eq:DropletThicknessEvolutionConvectionDiscrete}.

\begin{figure}[h!]
\includegraphics[width=0.98\textwidth]{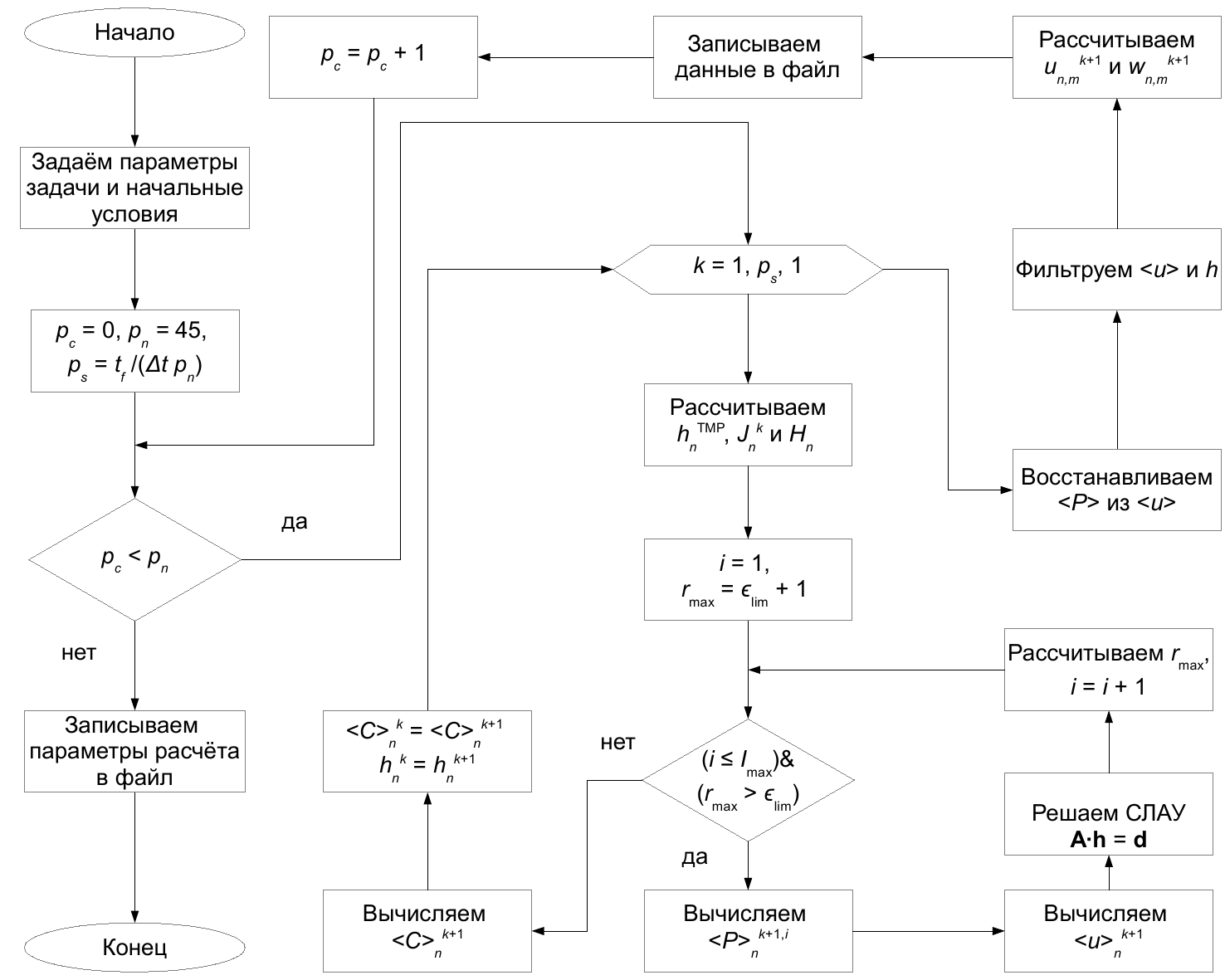}
\caption{Алгоритм численного расчета}\label{fig:NumericalAlgorithm}
\smallskip \footnotesize
[Figure~\ref{fig:NumericalAlgorithm}.
Numerical calculation algorithm]
\end{figure}

После того, как итерации завершились, полученные значения $h_n^{k+1}$, $\langle u\rangle_{n}^{k+1}$ и $\langle P\rangle_{n}^{k+1}$ используются для нахождения $\langle C \rangle_n^{k+1}$, $u_{n,m}^{k+1}$ и $w_{n,m}^{k+1}$ через явные зависимости~\eqref{eq:ConvectionDiffusionEquationDiscrete}, \eqref{eq:uVelocityDiscrete} и \eqref{eq:wVelocityDiscrete}. Но предварительно необходимо подавить пилообразные осцилляции, возникающие при расчёте давления. Этот вид осцилляций не приводит к <<разрушению>> численного счёта со временем, но поле скорости потока жидкости получится некорректным. Чтобы разрешить эту проблему, сделаем следующее. Выражение~\eqref{eq:AveragedRadialVelocityDimensionless} запишем как $$\frac{\partial \langle P\rangle}{\partial r} = - 3\frac{\langle u\rangle \eta}{h^2},$$ проинтегрируем по $r$, $$\left. \langle P\rangle \right|_r - \left. \langle P\rangle \right|_0 = -3 \int \limits_{0}^{r} \frac{\langle u\rangle \eta}{h^2}\,dr.$$ Слагаемое с интегралом аппроксимируем методом прямоугольников, получаем формулу для восстановления давления
\begin{equation}\label{eq:pressureRecovery}
	\langle P\rangle_n^\mathrm{repair} = -3 \Delta r \sum\limits_{n=0}^{N-1} \frac{\langle u\rangle \eta}{h^2} +  \langle P\rangle_0^\mathrm{repair}.
\end{equation}
 Учитывая пилообразную форму осцилляций, $\langle P\rangle_0^\mathrm{repair}$ можно аппроксимировать следующим образом, $\langle P\rangle_0^\mathrm{repair} \approx \left(\langle P\rangle_1 + \langle P\rangle_2 \right)/ 2$. Со временем такие пилообразные осцилляция могут передавться усредненной скорости $\langle u\rangle$ и профилю капли $h$. Здесь эта проблема решена за счет использования сглаживающего фильтра
\begin{equation}\label{eq:smoothingFilter}
	f_n^\mathrm{filt} = \beta f_{n+1} + \gamma f_{n} + \delta f_{n-1}.
\end{equation}
 Параметры подобраны эмпирическим путём, $\beta = \delta = 0.25$ и $\gamma = 0.5$. В отличие от~\cite{ParkY2019} используемый здесь фильтр прост в реализации, занимает в разы меньше памяти и вычислительных ресурсов. Кроме того, количество итераций фильтра из~\cite{ParkY2019} заранее неизвесно. В нашем случае достаточно одного прохода по узлам $n=1..N-1$. Узлы $n=0$ и $n=N$ рассчитываются из граничных условий.

Блок-схема с алгоритмом расчёта представлена на рис.~\ref{fig:NumericalAlgorithm}. В ней используются следующие дополнительные обозначения: $p_c$~--- счётчик временных периодов, $p_n$~--- количество временных периодов, $p_s$~--- размер временного периода во временных шагах и $H_n = H_a (x_n)$. Записи вида $f_n^k = f_n^{k+1}$ следует трактовать как векторные операции. С кодом программы на языке С++ можно ознакомиться \href{https://github.com/kolegovk/Suppression-of-sawtooth-oscillations.git}{здесь}.

Для сравнения численный расчет также выполнялся в коммерческом пакете Maple~ 18 с помощью модуля pdsolve с параметром method=$\Theta[1]$ (неявная разностная схема). Для того, чтобы pdsolve справился с задачей, систему уравнений пришлось переписать в другом виде. Вместо~\eqref{eq:AveragedRadialVelocityDimensionless} и \eqref{eq:CapillaryPressureDimensionless} использовались уравнения
\begin{equation}\label{eq:AuxiliaryEquationForMaple}
   \psi = - \frac{1}{\mathrm{Ca}} \frac{\partial h}{\partial r},
\end{equation}
\begin{equation}\label{eq:AveragedRadialVelocityForMaple}
\left\langle u\right\rangle = -H_a(x)\frac{h^2}{3 \eta} \frac{\partial}{\partial r}\left( \frac{1}{r} \frac{\partial (r \psi)}{\partial r} \right),
\end{equation}
где $\psi$~--- вспомогательная функция. Причём модуль pdsolve не требует граничных условий для~\eqref{eq:AuxiliaryEquationForMaple}. Скорее всего, модуль определяет их автоматически из заданных условий для $h$. Для уравнения~\eqref{eq:AveragedRadialVelocityForMaple} на границах задается равенство скорости нулю (см. раздел~\ref{subsec:IBC}). Таким образом, в Maple решалась система уравнений~\eqref{eq:DropletThicknessEvolutionDimensionless}, \eqref{eq:ConvectionDiffusionEquationDimensionless}, \eqref{eq:AuxiliaryEquationForMaple} и \eqref{eq:AveragedRadialVelocityForMaple}.

\Section{Результаты и обсуждение}

\begin{figure}
	
	\noindent
	\begin{minipage}[h]{0.5\textwidth}
		\centering  \scriptsize \sl
		\includegraphics[width=0.98\textwidth]{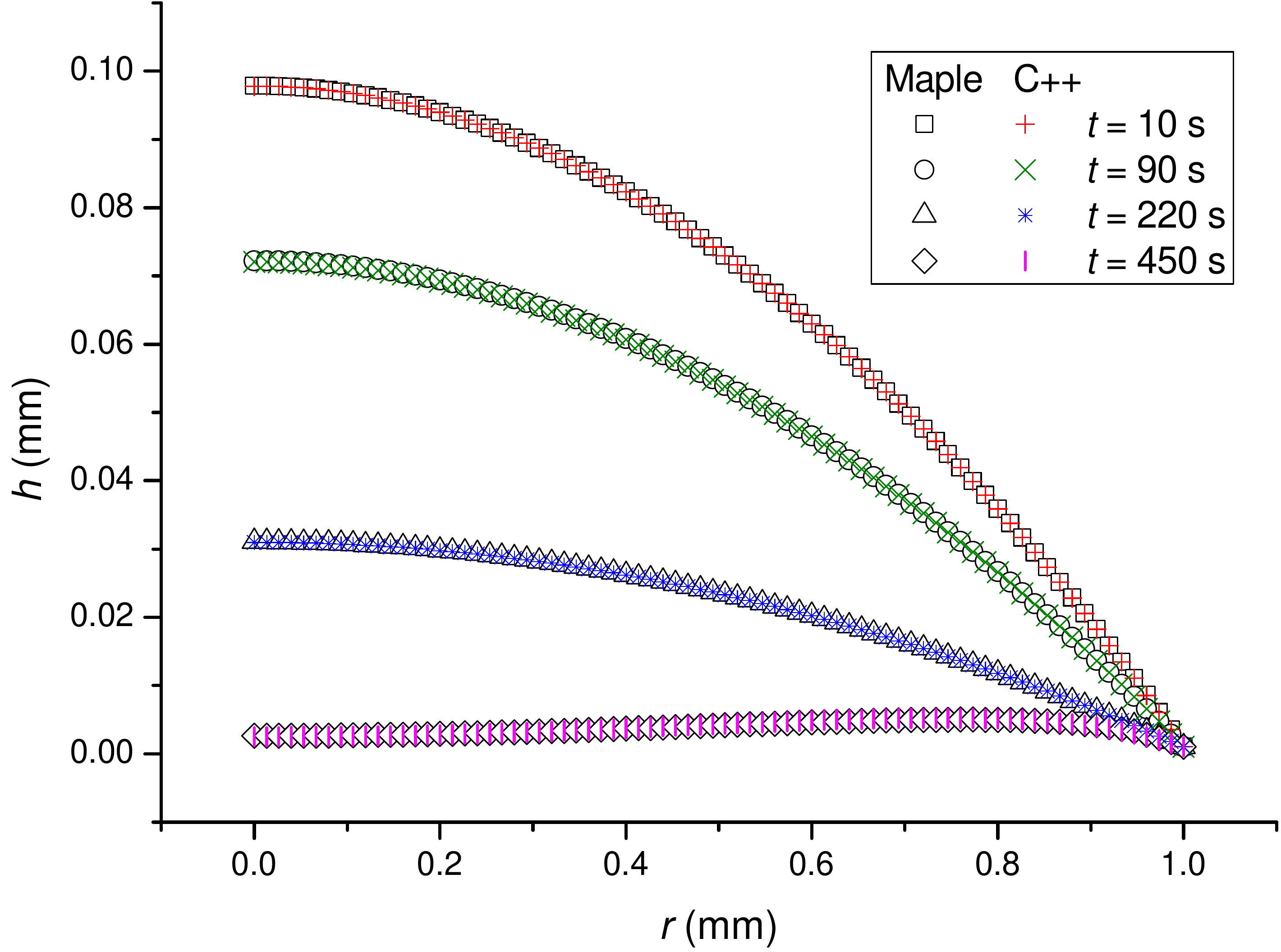}\\
		a
	\end{minipage}
	\begin{minipage}[h]{0.5\textwidth}
		\centering  \scriptsize \sl
		\includegraphics[width=0.98\textwidth]{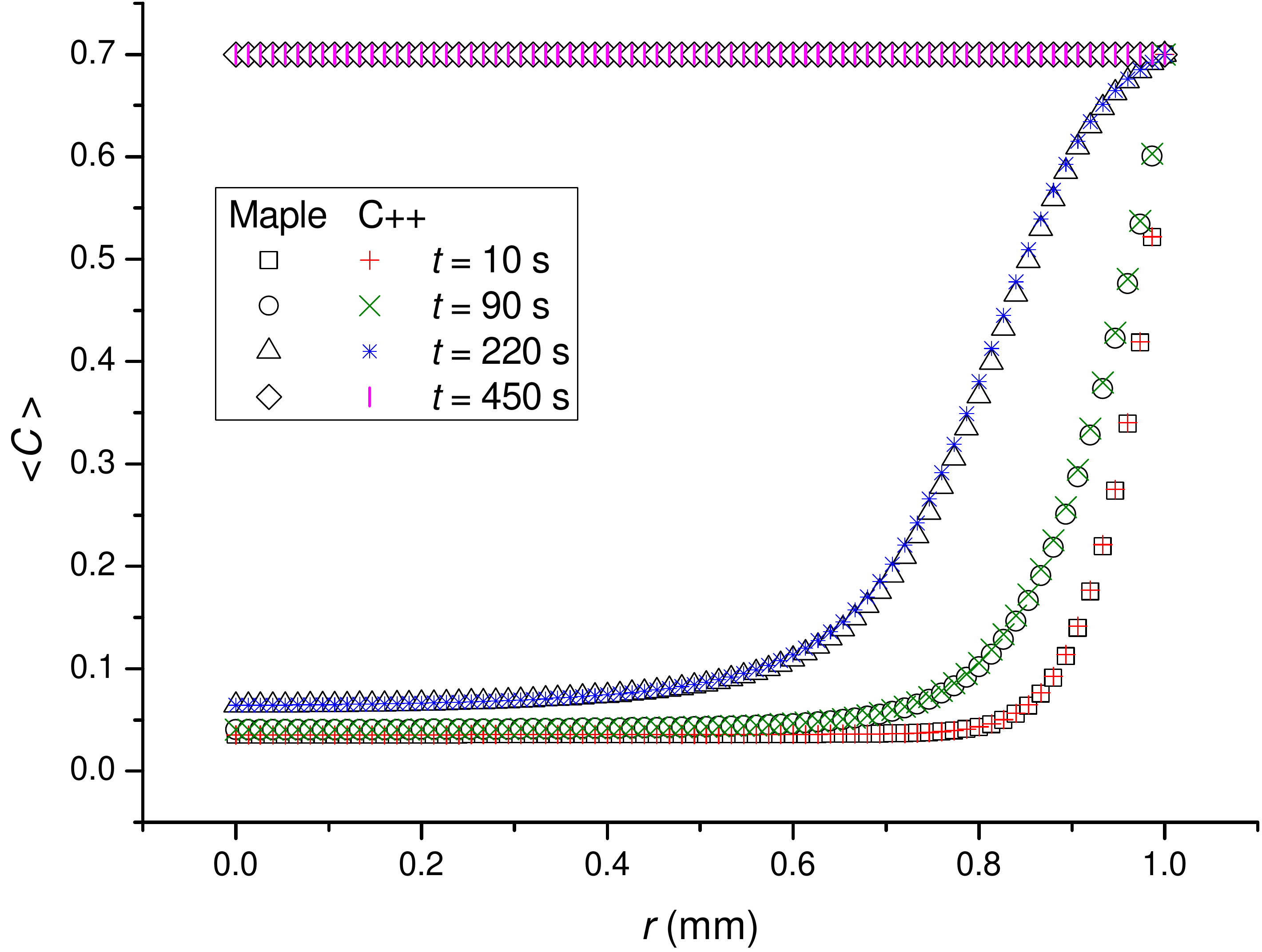}\\
		b
	\end{minipage}
	
	\vspace{2mm}
	
	\noindent
	\begin{minipage}[h]{0.5\textwidth}
		\centering  \scriptsize \sl
		\includegraphics[width=0.98\textwidth]{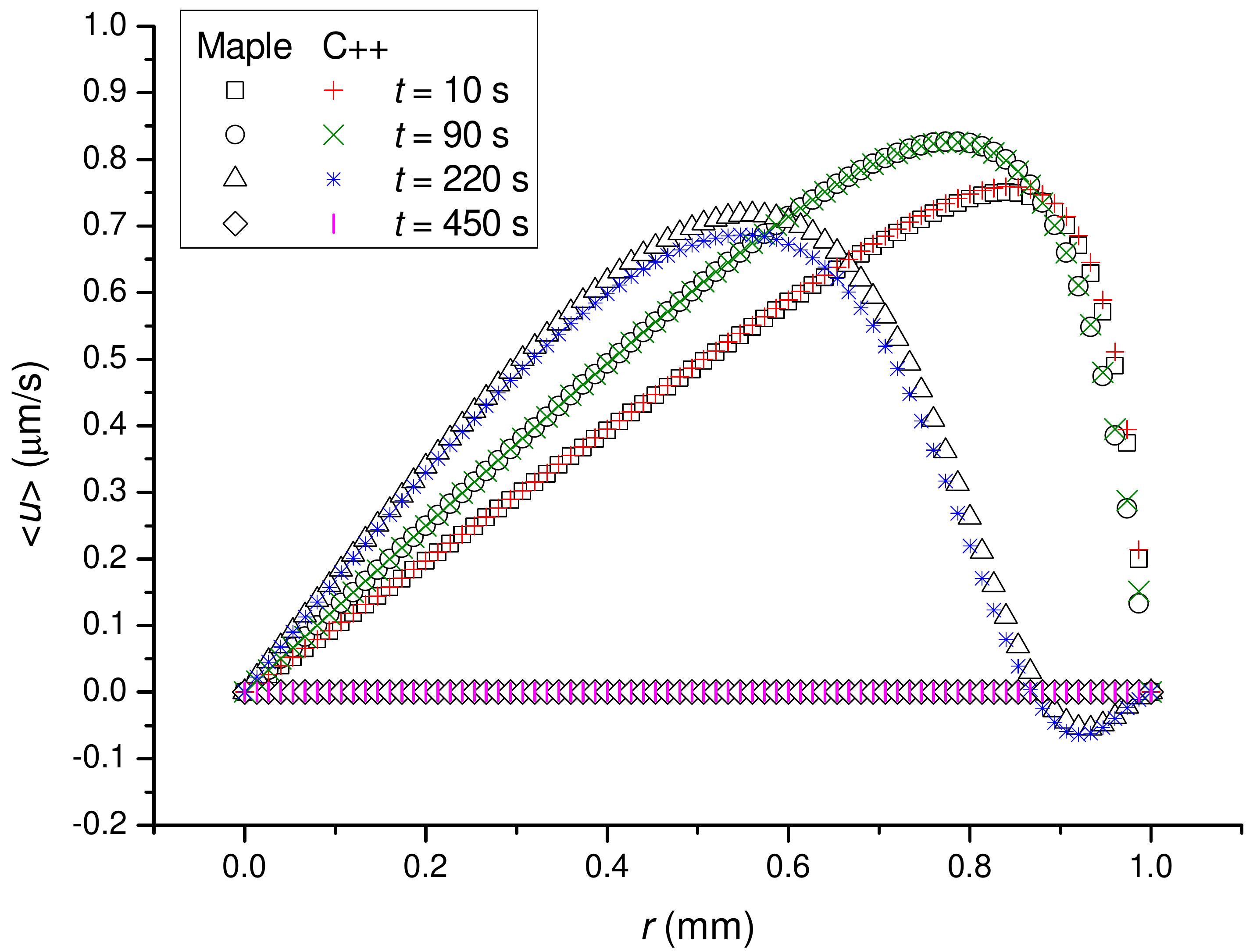}\\
		c
	\end{minipage}
	\hfill
	\begin{minipage}[h]{0.48\textwidth}
		\caption{Сравнение результатов расчётов в двух программах для нескольких последовательных моментов времени: (a) высота капли, (b) массовая доля растворённого или взвешенного вещества и (c) радиальная скорость потока, усреднённая по толщине жидкого слоя \label{comparisonCvsMaple}}
	\end{minipage}
	
	\smallskip \footnotesize
	
	[Figure~\ref{comparisonCvsMaple}.
	Comparison of the calculation results in two programs for several consecutive time points: (a) drop height, (b) mass fraction of dissolved or suspended matter, and (c) radial flow velocity averaged over the thickness of the liquid layer]
\end{figure}

Результаты расчетов средствами двух программ сравниваются на рис.~\ref{comparisonCvsMaple}. На большинстве графиков расхождение практически не заметно. Максимальное отличие результата, полученного с помощью программы, написанной на С++, от результата, полученного с помощью модуля pdsolve в Maple 18, наблюдается для $\langle u\rangle$ на времени $t =$ 220~с согласно рис.~\ref{comparisonCvsMaple}c. В абсолютных величинах наибольшее расхождение соответсвует точке $r\approx$ 0.75~мм и составляет примерно 0.046~мкм/с, что в относительных величинах около 10$\%$.

Расчет выполнялся для $N=75$ в двух программах. Значения временных шагов для разработанной программы и для модуля pdsolve брались разные, так как в них используются разные алгоритмы расчёта, соответственно и требования к временным шагам накладываются тоже разные. В Maple 18 для численного решения СЛАУ используется итерационный метод Ньютона, но технические детали реализации закрыты для пользователя, есть лишь краткая справка по разностным схемам. Скорость вычислений с относительно малым временным шагом конечно же ниже, но это частично компенсируется тем, что C++ является компилируемым языком, а язык Maple~--- интерпретируемый. Для расчетов в Maple использовалось значение $\Delta t =$ 0.1~с, в разработанной программе~--- $\Delta t =10^{-6}$~с. Кроме того, для программы на Си++ выполнены тестовые расчёты для $N = 25$ и $N = 50$ с целью проверки сходимости по сетке (во всех расчётах $M=N$). Для $N = 25$ приближённо подобрано максимальное значение временного шага $\Delta t =10^{-4}$~с, при котором расчет не <<разрушается>>. Значению $N = 50$ соответствует $\Delta t =5\cdot 10^{-6}$~с. Таким образом, небольшое уменьшение $\Delta r= R/N$ приводит к тому, что нужно значительно уменьшать временной шаг. Эти наблюдения в вычислительных экспериментах косвенно свидетельствуют о том, что предложенная разностная схема является условно устойчивой. Длительность расчёта в Maple составила около 10 мин для $N=75$ (объём занимаемой памяти $\approx 217$~Мб). Время расчёта в программе на С++ примерно составило: 10 мин для $N = 25$,  5 ч для $N = 50$ и 29 ч для $N=75$ (объём занимаемой памяти $\approx 3$~Мб) в связи с необходимостью уменьшать временной шаг. Расчеты проводились на компьютере с ЦП Intel Xeon CPU E3-1230 v3 c максимальной частотой 3.6 ГГц под управлением ОС Windows 10. Для программы на С++ использовался компилятор, встроенный в Visual Studio 2022 Community Edition.

Проверка закона сохранения массы для растворённого или взвешенного вещества для $N=75$ показала погрешность около $0.27\%$, что экспериментально подтверждает свойство консервативности в предложенной разностной схеме. Для этого были рассчитаны и сопоставлены значения $$\sum\limits_{n=0}^{N} h_n \langle C \rangle_n$$ на моментах времени $t=0$ и $t=t_f$. Для расчётов в Maple также проводилась  проверка закона сохранения массы. Выполнены тестовые расчёты и определены погрешности для разных временных шагов в диапазоне, для которого численное решение итерационно сходится:  $\Delta t =$ 0.01~с~--- $0.43\%$, $\Delta t =$ 0.1~с~--- $0.1\%$, $\Delta t =$ 1~с~--- $3.11\%$  и для $\Delta t =$ 10~с~--- $32.54\%$. Минимальная погрешность в $0.1\%$ соответствует временному шагу $\Delta t =$ 0.1~с, который и был выбран для расчётов в Maple.

\begin{figure}
	
	\noindent
	\begin{minipage}[h]{0.5\textwidth}
		\centering  \scriptsize \sl
		\includegraphics[width=0.98\textwidth]{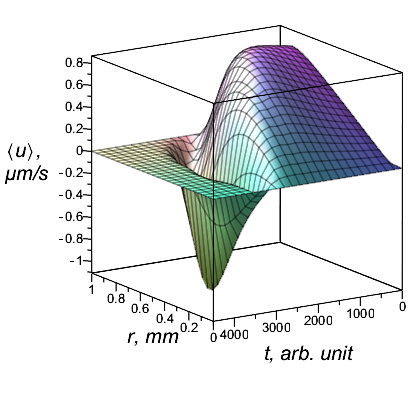}\\
		a
	\end{minipage}
	\begin{minipage}[h]{0.5\textwidth}
		\centering  \scriptsize \sl
		\includegraphics[width=0.98\textwidth]{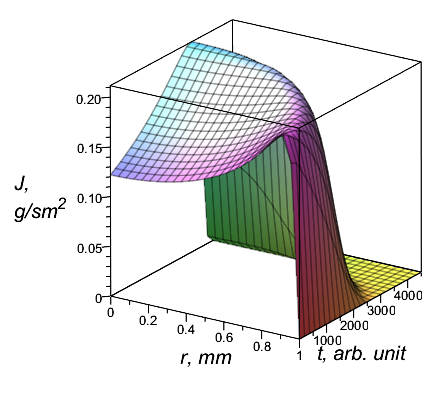}\\
		b
	\end{minipage}
	\caption{Результаты расчёта в Maple: пространственно-временная зависимость (a) усреднённой по толщине жидкого слоя скорости радиального потока и (b) плотности потока пара \label{fig:RadialVelAndEvapFluxDensVsTime}}
	\smallskip \footnotesize
	[Figure~\ref{fig:RadialVelAndEvapFluxDensVsTime}.
	The results of the calculation in Maple: the spatial-temporal dependence of (a) the radial flow velocity averaged over the thickness of the liquid layer and (b) the vapor flux density]
\end{figure}

Толщина жидкого слоя с течением времени уменьшается (рис.~\ref{comparisonCvsMaple}a). Из-за пространственно неравномерного испарения происходит отклонение формы капли от равновесного состояния, возникает градиент капиллярного давления, что порождает капиллярный поток (рис.~\ref{comparisonCvsMaple}c), переносящий вещество к периферии капли. В связи с этим массовая доля вещества, содержащегося в жидкости, увеличивается вблизи трёхфазной границы со временем (рис.~\ref{comparisonCvsMaple}b). Под конец процесса массовая доля $\langle C \rangle$  распределена равномерно, так как достигнуто критическое значение $C_g$, при котором произошёл фазовый переход. Дальнейшее испарение жидкости из твёрдой фазы здесь не рассматривается.

Финальная расчётная толщина слоя $h$ является почти равномерной в пространстве. Вблизи $r\approx$ 0.9~мм наблюдается незначительное возвышение, которое связано с эффектом кофейных колец~\cite{Deegan1997}. Возникающий под конец процесса обратный поток жидкости, направленный в центральную область, смазывает этот эффект. Это можно использовать для улучшения существующих методов подавления эффекта кофейных колец (см. ссылки в~\cite{Zang2019}).

\begin{figure}
	\noindent
	\begin{minipage}[h]{0.5\textwidth}
		\centering  \scriptsize \sl
		\includegraphics[width=0.98\textwidth]{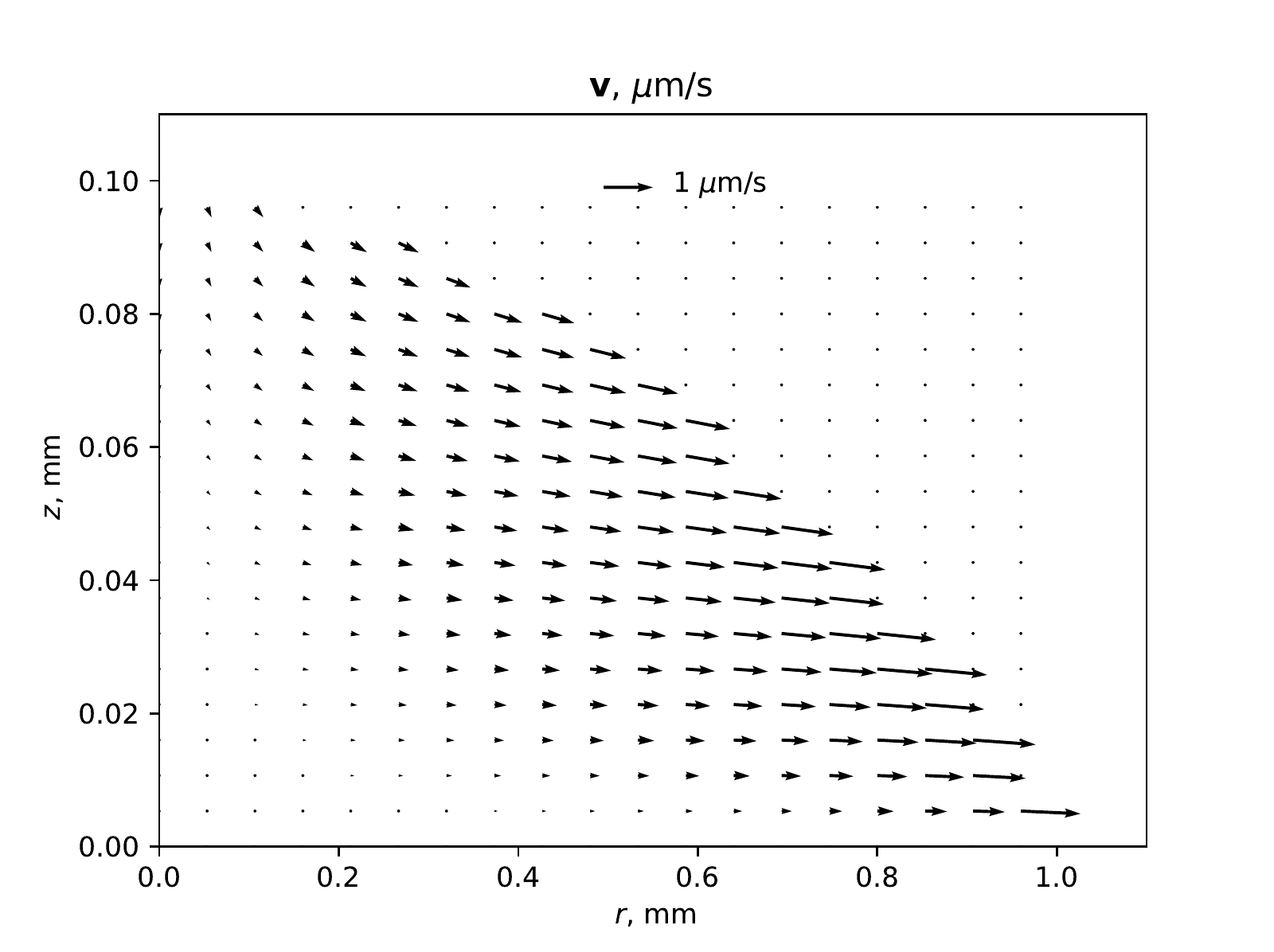}\\
		a
	\end{minipage}
	\begin{minipage}[h]{0.5\textwidth}
		\centering  \scriptsize \sl
		\includegraphics[width=0.98\textwidth]{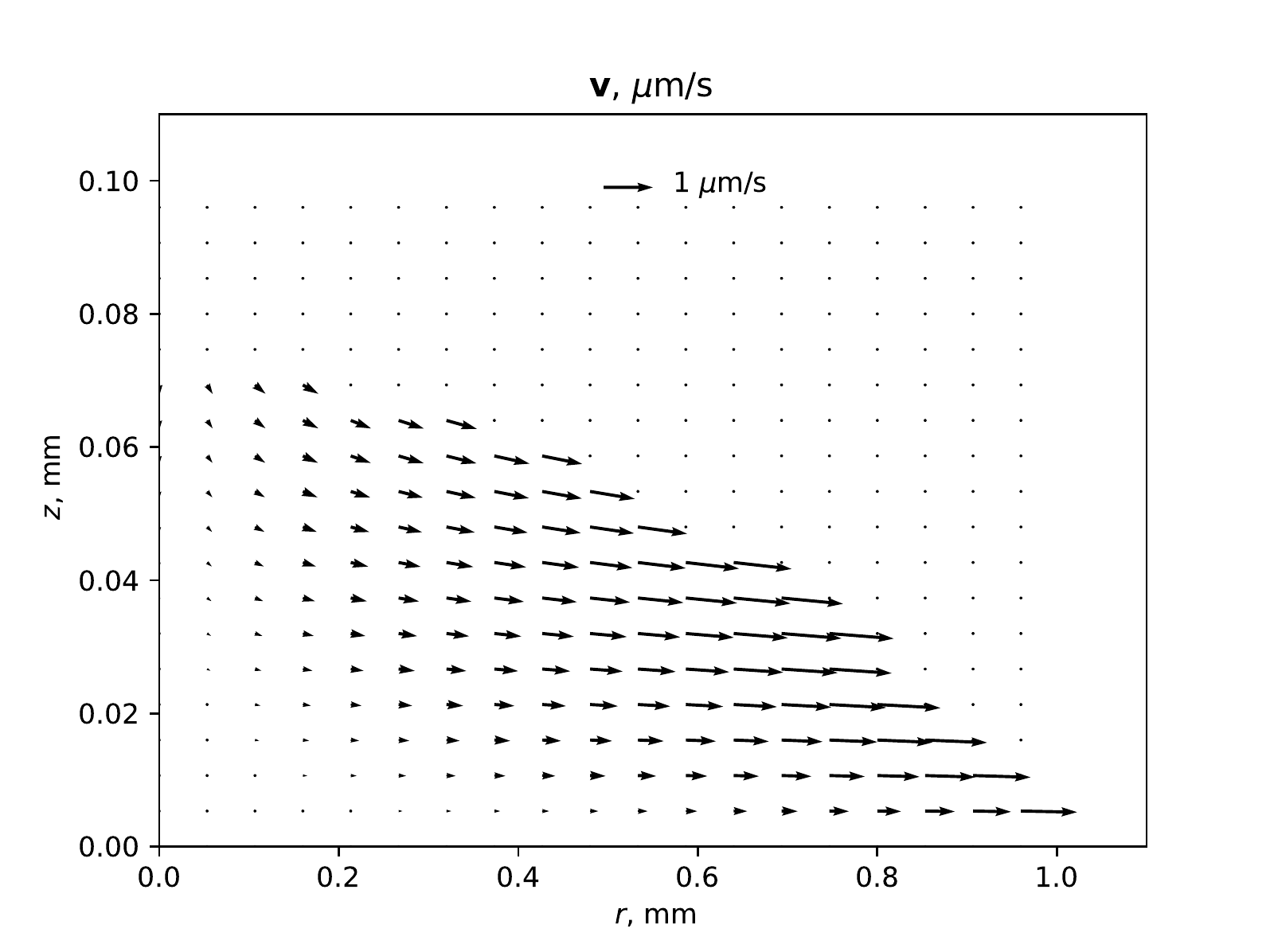}\\
		b
	\end{minipage}
	\vspace{2mm}
	\noindent
	\begin{minipage}[h]{0.5\textwidth}
		\centering  \scriptsize \sl
		\includegraphics[width=0.98\textwidth]{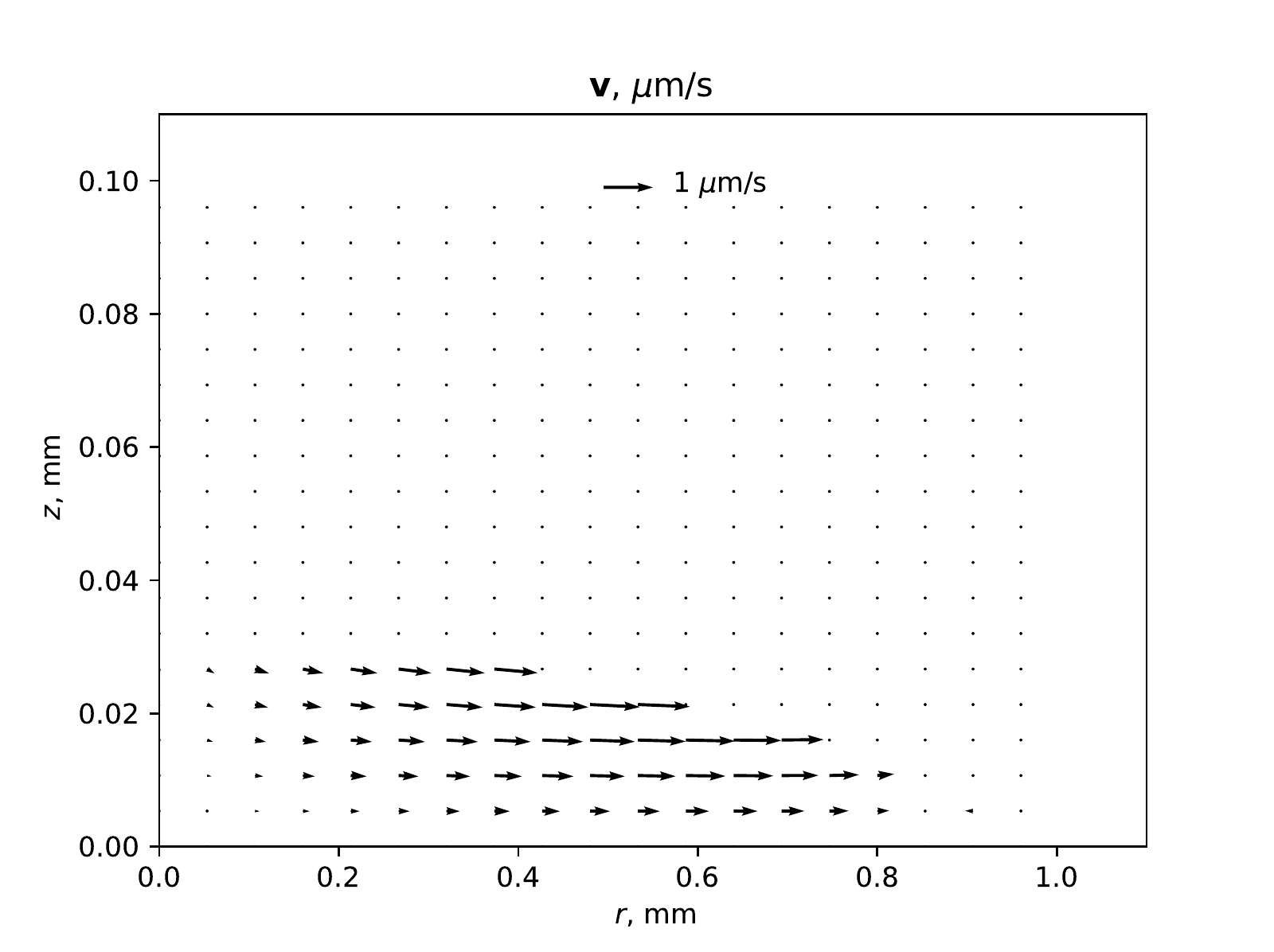}\\
		c
	\end{minipage}
	\begin{minipage}[h]{0.5\textwidth}
		\centering  \scriptsize \sl
		\includegraphics[width=0.98\textwidth]{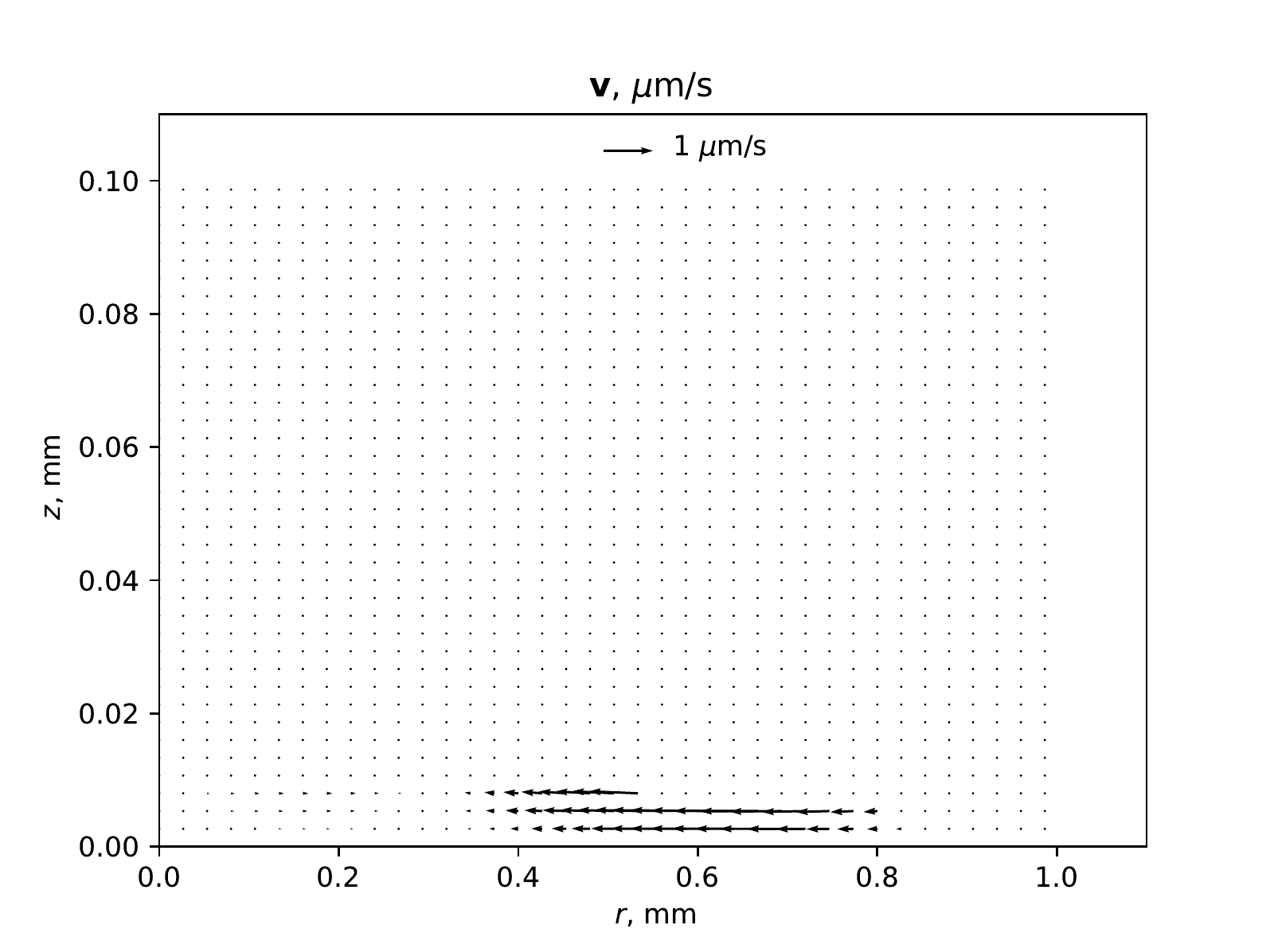}\\
		d
	\end{minipage}
	\caption{Векторное поле скорости потока жидкости по результатам, полученным с использованием программы, написанной на C++, для времени (a) $t=$ 10~с, (b) $t=$ 90~с, (c) $t=$ 220~с и (d) $t=$ 300~с \label{fig:velocityVectorFieldOfFluidFlow}}
	\smallskip \footnotesize
	[Figure~\ref{fig:velocityVectorFieldOfFluidFlow}.
	Velocity vector field of fluid flow according results obtained with using the program written in C++ for the time (a) $t=$ 10~s, (b) $t=$ 90~s, (c) $t=$ 220~s, and (d) $t=$ 300~s]
\end{figure}

Также стоит заметить, что поначалу усреднённая радиальная скорость потока с течением времени увеличивается в связи с увеличением градиента капиллярного давления в процессе неравномерого испарения, преобладающего вблизи периферии. Затем в области роста массовой доли вещества, приводящего к увеличению вязкости, поток начинает замедляться. Вблизи трёхфазной границы появляется противоток  (рис.~\ref{comparisonCvsMaple}c для $t =$ 220 с), что ранее наблюдалось в эксперименте~\cite{Bodiguel2010}. Далее по времени направление потока постепенно меняется во всей области (рис.~\ref{fig:RadialVelAndEvapFluxDensVsTime}a). Это связано с тем, что в начале процесса плотность потока пара преобладает в районе периферии, но со временем $J$ начинает увеличиваться в центральной области из-за понижения $h$ и уменьшаться вблизи контактной линии из-за роста $\langle C \rangle$ (рис.~\ref{fig:RadialVelAndEvapFluxDensVsTime}b). Когда во всей области достигается значение $\langle C \rangle \approx C_g$, модель предсказывает отсутсвие потока и испарения. Поле скорости потока для различных моментов времени представлено на рис.~\ref{fig:velocityVectorFieldOfFluidFlow}. В отличие от результатов расчётов~\cite{Fischer2002} здесь на поздних временных этапах наблюдается изменение направления потока. Это связано с тем, что в текущей работе учитывается зависимость $J$ и $\eta$ от массовой доли $\langle C \rangle$. Кроме того, здесь при заданных параметрах градиент плотности потока пара с течением времени меняет знак. Технические детали численного решения в~\cite{Fischer2002} не раскрыты. В~\cite{Tarasevich2011,Kolegov2018344,ParkY2019} поле скорости потока не рассчитывалось.

\begin{figure}
	
	\noindent
	\begin{minipage}[h]{0.5\textwidth}
		\centering  \scriptsize \sl
		\includegraphics[width=0.98\textwidth]{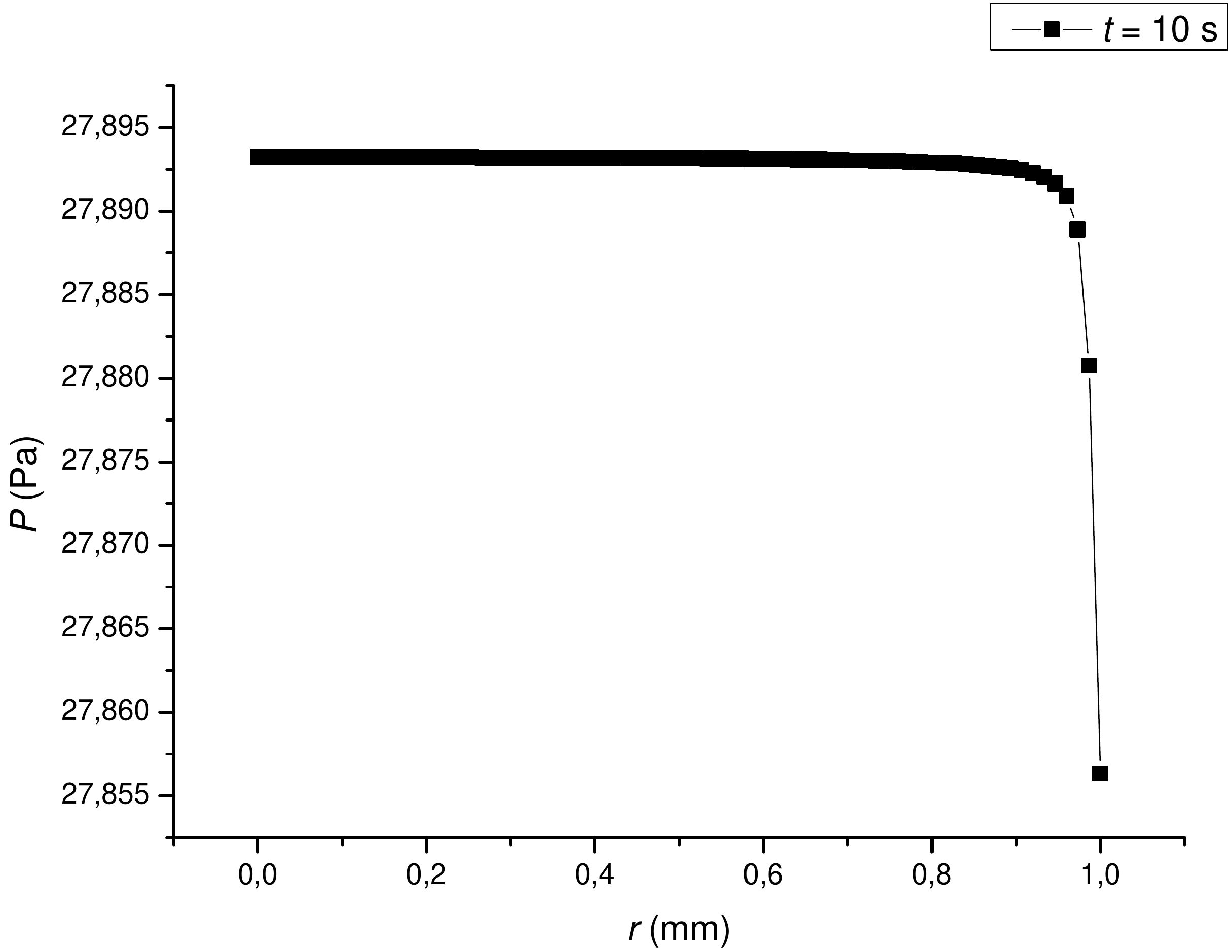}\\
		a
	\end{minipage}
	\begin{minipage}[h]{0.5\textwidth}
		\centering  \scriptsize \sl
		\includegraphics[width=0.98\textwidth]{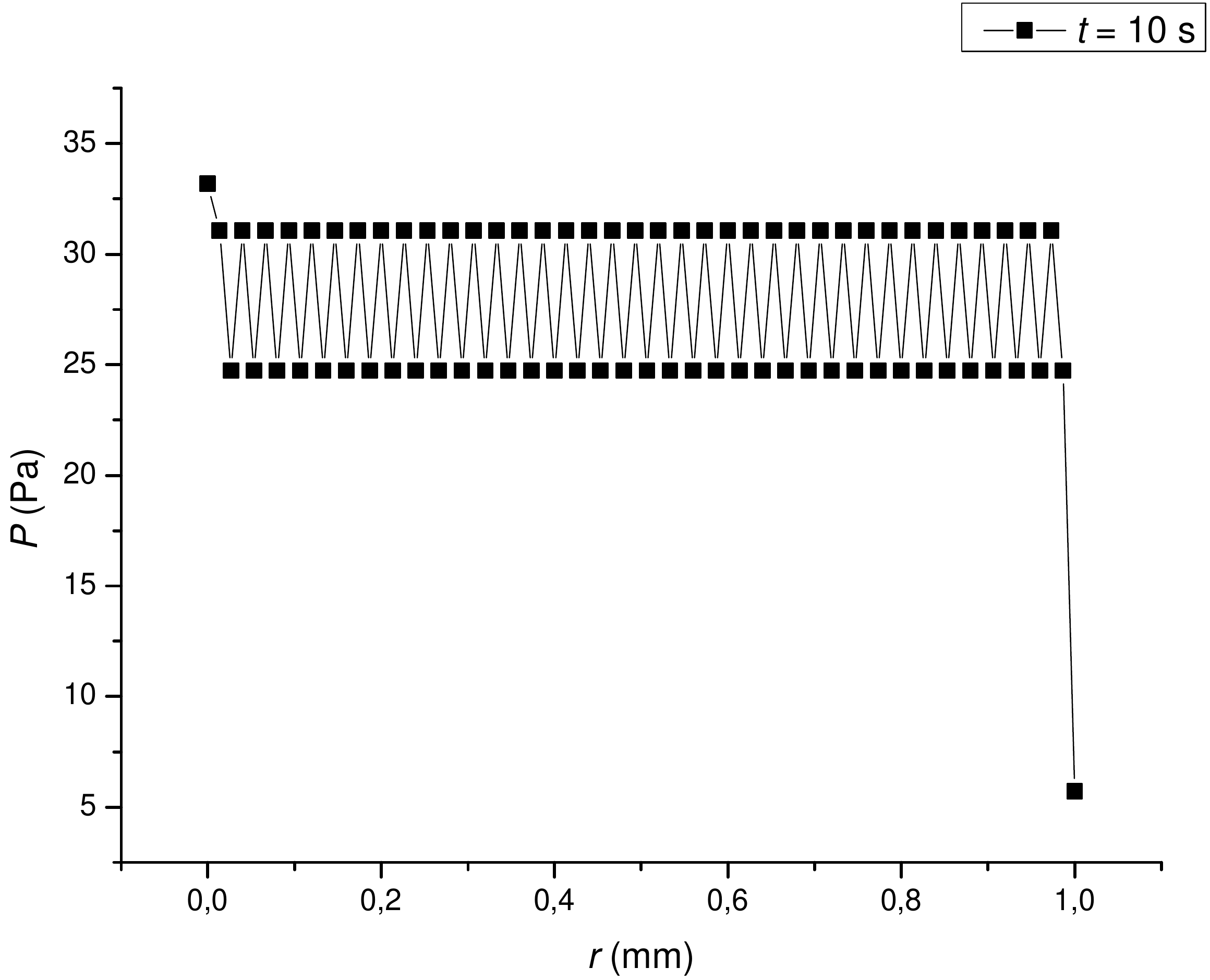}\\
		b
	\end{minipage}
	\caption{Результаты расчёта в программе, написанной на С++: (a) капиллярное давление и (b) возможные пилообразные осцилляции \label{fig:pressureOscillation}}
	\smallskip \footnotesize
	[Figure~\ref{fig:pressureOscillation}.
	The results of the calculation in program written with C++: (a) capillary pressure and (b) possible sawtooth oscillations]
\end{figure}

Здесь поле скорости потока удалось рассчиать по формулам~\eqref{eq:uVelocityDiscrete} и \eqref{eq:wVelocityDiscrete} лишь после того, как была решена проблема с пилообразными осцилляциями, возникающими при расчёте капиллярного давления и постепенно с временными шагами накапливающимися в $\langle u\rangle$ и $h$. Для этого использовались формулы~\eqref{eq:pressureRecovery} и \eqref{eq:smoothingFilter}. Восстановленное из скорости потока $\langle u\rangle$ по формуле~\eqref{eq:pressureRecovery} капиллярное давление представлено на рис.~\ref{fig:pressureOscillation}a. Если такое восстановление не проводить, то получаем осцилляции, представленные на рис.~\ref{fig:pressureOscillation}b. Особенность этих осцилляций заключается в стабильности на временных шагах, то есть они не приводят к полному <<разрушению>> расчёта. Пилообразные осцилляции теоретически изучаются в~\cite{Dorodnitsyn2016}, но здесь предложено практическое решение проблемы на конкретном примере.

\Section[N]{Заключение}
Для решения практических задач в области испарительной самосборки и испарительной литографии требуется разработка математических моделей, численных методов и комплекса программ. При разработке необходимо учитывать особенности таких задач, ведь сопутствующие процессы весьма сложны. Вязкость раствора, градиент плотности потока пара и локальная кривизна свободной поверхности влияют на структуру потока в сидячей кале. В работе показан случай, когда направление потока жидкости может меняться со временем в процессе испарения даже в отсутствии эффекта Марангони и какого-либо внешнего воздействия на систему. Это может приводить к замедлению выноса вещества на периферию, что в результате будет способствовать формированию более или менее равномерного осадка по всей площади контакта капли с подложкой. Данное наблюдение полезно для совершенствования методов подавления кольцевых осадков, связанных с эффектом кофейных колец и нежелательных для некоторых приложений, как, например, струйная печать или нанесение покрытий. Моделирование гидродинамических потоков важно, так как это один из основных механизмов, влияющих на массоперенос растворённого или взвешенного вещества. При численной реализации подобных задач может возникнуть проблема, связанная с возникновением пилообразных осцилляций, устойчивых на временных шагах, но создающих проблему в расчёте поля скорости потока жидкости. Здесь предложены практические рецепты, позволяющие эффективно бороться с такими осцилляциями при разработке программного обеспечения для вычислительной гидродинамики.

\AdditionalContent{Конкурирующие интересы}{Заявляю, что в отношении авторства и публикации этой статьи конфликта интересов не имею.}
\AdditionalContent{Финансирование}{Часть работы, относящаяся к разработке математической модели, выполнена при поддержке программы развития университета <<Приоритет 2030>> проект №~122112500011-0 <<Природовдохновленные оптические технологии>>, Центра природовдохновленного инжиниринга в ТюмГУ. Исследование, связанное с разработкой численного метода и программы для ЭВМ, выполнено за счет гранта
Российского научного фонда №~22-79-10216 в АГУ им. В.Н.~Татищева, \href{https://rscf.ru/project/22-79-10216/}{https://rscf.ru/project/22-79-10216/}.}
\AdditionalContent{Благодарность}{Автор выражает благодарность профессору, д.ф.-м.н. Тарасевичу~Ю.Ю. за полезные рекомендации по оформлению графического материала.}

\bibliographystyle{ugost2008}

\bibliography{biblio}

\newpage

\makeentitle

\vspace{-5mm}

\AdditionalContent{Competing interests}{I have no competing interests.}

\AdditionalContent{Funding}{Part of the work related to the development of a mathematical model was carried out with the support of the University development program ``Priority 2030'' project 122112500011-0 ``Bioinspired optical engineering'' of Centre for Nature-Inspired Engineering at University of Tyumen. The work related to the development of the numerical method and the computer program is supported by the grant 22-79-10216 from the Russian Science Foundation at ASU named after V.N. Tatishchev (\href{https://rscf.ru/en/project/22-79-10216/}{https://rscf.ru/en/pro\\ject/22-79-10216/}).}

\AdditionalContent{Acknowledgments}{The author is grateful to Prof. Yu.~Yu.~Tarasevich for useful recommendations on the design of graphic material.}

\end{document}